\documentclass[preprint]{aastex}
\usepackage[]{natbib}
\usepackage{graphics}
 
\newcommand{\etal}{et al.}  \newcommand{\per}{\ensuremath{^{-1}}}
\newcommand{\persq}{\ensuremath{^{-2}}}
\newcommand{\percu}{\ensuremath{^{-3}}}

\newcommand{\hal}{H\ensuremath{\alpha}}
 \newcommand{\hst}{\emph{HST}}
\newcommand{\msun}{M\ensuremath{_{\odot}}}
 \newcommand{\kms}{km
s\ensuremath{^{-1}}} 

\newcommand{\lam}[1]{\ensuremath{\lambda}#1}

\newcommand{\ml}{\ensuremath{\Upsilon}}
\newcommand{\inc}{\ensuremath{i}}
\newcommand{\mbh}{\ensuremath{M_\mathrm{\bullet}}}
\newcommand{\subsamp}{\emph{s}} \newcommand{\hn}{\hal+[\ion{N}{2}]}
\newcommand{\chisq}{\ensuremath{\chi^2}}

\newcommand{\chisqdof}{\ensuremath{\chi^2/d}}

\newcommand{\vstar}{\ensuremath{v_\star}}
\newcommand{\vd}{\ensuremath{v_d}}
\newcommand{\rfit}{\ensuremath{r_\mathrm{fit}}}
\newcommand{\rhalf}{\ensuremath{r_{1/2}}}
\newcommand{\vsys}{\ensuremath{v_\mathrm{sys}}}
\newcommand{\lsun}{\ensuremath{L_\odot}}
\newcommand{\mbul}{\ensuremath{M_\mathrm{bulge}}}
\newcommand{\mlsun}{\ensuremath{\ml_\odot}}
 
\newcommand{\sigmalsf}{\ensuremath{\sigma_\mathrm{LSF}}}
\newcommand{\sigmath}{\ensuremath{\sigma_\mathrm{th}}}

\slugcomment{}
\shorttitle{BLACK HOLE IN NGC 3245} 
\shortauthors{BARTH ET AL.}

\begin{document} 

\title{Evidence for a Supermassive Black Hole in the S0 Galaxy NGC
3245\footnotemark[1]}

\footnotetext[1]{Based on observations with the NASA/ESA \emph{Hubble
Space Telescope} obtained at STScI, which is operated by AURA, Inc.,
under NASA contract NAS5-26555.}

\author{Aaron J. Barth\altaffilmark{2}, Marc Sarzi\altaffilmark{3,4},
Hans-Walter Rix\altaffilmark{4}, Luis C. Ho\altaffilmark{5}, Alexei
V. Filippenko\altaffilmark{6}, and Wallace
L. W. Sargent\altaffilmark{7}}

\altaffiltext{2}{Harvard-Smithsonian Center for Astrophysics, 60 Garden Street,
Cambridge, MA 02138}

\altaffiltext{3}{Dipartimento di Astronomia, Universit\`a di Padova, Vicolo
dell'Osservatorio 5, I-35122, Italy}

\altaffiltext{4}{Max-Planck-Institut f\"ur Astronomie, K\"onigstuhl 17,
Heidelberg D-69117, Germany}

\altaffiltext{5}{The Observatories of the Carnegie Institution of Washington,
813 Santa Barbara Street, Pasadena, CA 91101}

\altaffiltext{6}{Department of Astronomy, University of California,
Berkeley, CA 94720-3411}

\altaffiltext{7}{Palomar Observatory, 105-24 Caltech, Pasadena, CA 91125}

\begin{abstract}

The S0 galaxy NGC 3245 contains a circumnuclear disk of ionized gas
and dust with a radius of 1\farcs1 (110 pc), making it an ideal target
for dynamical studies with the \emph{Hubble Space Telescope} (\hst).
We have obtained spectra of the nuclear disk with the Space Telescope
Imaging Spectrograph, using an 0\farcs2-wide slit at five parallel
positions.  Measurements of the \hal\ and [\ion{N}{2}] emission lines
are used to map out the kinematic structure of the disk in
unprecedented detail.  The data reveal a rotational velocity field
with a steep velocity gradient across the innermost 0\farcs4.  We
construct dynamical models for a thin gas disk in circular rotation,
using \hst\ optical images to map out the gravitational potential due
to stars.  Our modeling code includes the blurring due to the
telescope point-spread function and the nonzero slit width, as well as
the instrumental shift in measured wavelength for light entering the
slit off-center, so as to simulate the data as closely as possible.
The \hn\ surface brightness measured from an \hst\ narrow-band image
is folded into the models, and we demonstrate that many of the
apparent small-scale irregularities in the observed velocity curves
are the result of the patchy distribution of emission-line surface
brightness.  Over most of the disk, the models are able to fit the
observed radial velocity curves closely, although there are localized
regions within the disk which appear to be kinematically disturbed
relative to the overall rotational pattern.  The velocity dispersion
of [\ion{N}{2}] \lam6584 rises from $\sigma\approx50$ \kms\ in the
outer disk to $\sim160$ \kms\ at the nucleus, and most of this
linewidth cannot be attributed to rotational or instrumental
broadening.  To account for the possible dynamical effect of the
intrinsic velocity dispersion in the gas, we also calculate models
which include a correction for asymmetric drift.  This correction
increases the derived black hole mass by 10\%, but leads to slightly
poorer fits to the data.  A central dark mass of
$(2.1\pm0.5)\times10^8$ \msun\ is required for the models to reproduce
the steep central velocity gradient.  This value for the central mass
is consistent with recently discovered correlations between black-hole
mass and bulge velocity dispersion.

\end{abstract}


\section{Introduction}

During the past few years, tremendous progress has been made in the
quest to detect supermassive black holes (BHs) in galactic nuclei.
Over 30 dynamical measurements of BH masses have now been performed,
most of them based on data from the \emph{Hubble Space Telescope}
(\hst) \citep[for reviews, see][]{kr95, ric98, ho99, kor00}.  These
measurements have confirmed and extended early suggestions
\citep{kr95} that BH mass is correlated with the mass of the bulge
component of the host galaxy, albeit with substantial scatter.  An
intriguing recent development has been the discovery that the BH mass
and bulge velocity dispersion are very tightly correlated \citep{fm00,
geb00, mf00}.  Models of bulge formation and the growth of ``seed''
BHs and quasar engines have provided theoretical frameworks for
understanding the origin of this correlation \citep{sr98, ost00, hk01,
agr01}.

One issue remaining to be resolved is the exact slope of the
correlation between BH mass and stellar velocity dispersion.
\citet{geb00} and \citet{fm00} find different values for the slope,
and the various theoretical scenarios make different predictions for
this quantity as well.  To better understand BH demographics, more
mass measurements are needed, and with the highest precision possible.
Several \hst\ surveys currently in progress will provide a wealth of
new data on BH masses, using both stellar-dynamical and gas-dynamical
measurement techniques.

Measurements of ionized gas kinematics in the environment of a BH can
provide the most straightforward means to determine the central mass,
provided that the gas is in circular rotation.  Using \hst\
spectroscopy, the gas-dynamical method has successfully been applied
to several early-type galaxies and a few spirals.  A handful of
objects were observed with the \hst\ Faint Object Spectrograph (FOS)
\citep{har94, ffj96, vv98, ff99, vk00}.  As a single-aperture
spectrograph, however, the FOS was an inefficient instrument for
mapping velocity fields.  The long-slit mode of the Faint Object
Camera (FOC) was used successfully in one case, to measure gas
kinematics in the nucleus of M87 \citep{mac97}.  Since the
installation of the Space Telescope Imaging Spectrograph (STIS), a
long-slit spectrograph, it has become possible to map out the velocity
structure of ionized gas disks very efficiently, as demonstrated by
\citet{bow98}, \citet{shi00}, and \citet{sar01}.

STIS observations provide significant advantages in comparison with
the previous generation of \hst\ spectrographs, and BH masses can
consequently be measured with greater precision.  The ability to
measure gas velocities at many positions simultaneously means that the
assumption of circular rotation can be tested much more thoroughly,
and the disk orientation can be constrained with kinematic data.  At
the same time, the improvement in data quality also makes it possible
to perform more detailed and accurate modeling of the observations,
which in turn leads to better constraints on the BH mass.  Recent
studies have also begun to examine the nuclear disk structure in more
detail, including the possibility that asymmetric drift may affect the
disk dynamics \citep{vk00}.

This paper presents an analysis of \hst\ observations of NGC 3245, as
part of a STIS program designed to study the dynamics of ionized gas
in nearby, low-luminosity active galactic nuclei (AGNs).  NGC 3245 is
an S0 galaxy with a mildly active nucleus classified as a
LINER/\ion{H}{2} ``transition'' type \citep{hfs97a}.  Ground-based
spectra displayed by \citet{hfs95} show double-peaked profiles of
\hal\ and [\ion{N}{2}] \lam6584, indicating spatially unresolved,
rapid rotation near the nucleus.  VLA observations reveal an
unresolved nuclear radio source, with no apparent jet
\citep{wh91}. NGC 3245 lies at a distance modulus of $31.60 \pm 0.20$
(a distance of 20.9 Mpc), as determined by measurements of
surface-brightness fluctuations \citep{ton01}.

The \hst\ observations, the basic measurements, and the deprojection
of the galaxy's surface-brightness profile are described in
\S\ref{wfpc2} and \S\ref{stis}.  In \S\ref{modeling} we give a
detailed description of the technique used to model the velocity
field, and \S\ref{results} describes the results of this analysis.  In
\S\ref{discussion} we discuss this work in the context of other BH
mass measurements and the correlations with host galaxy properties.

\section{WFPC2 Imaging}
\label{wfpc2}

\subsection{Observations}

\addtocounter{footnote}{1}

\begin{figure}
\begin{center}
\includegraphics{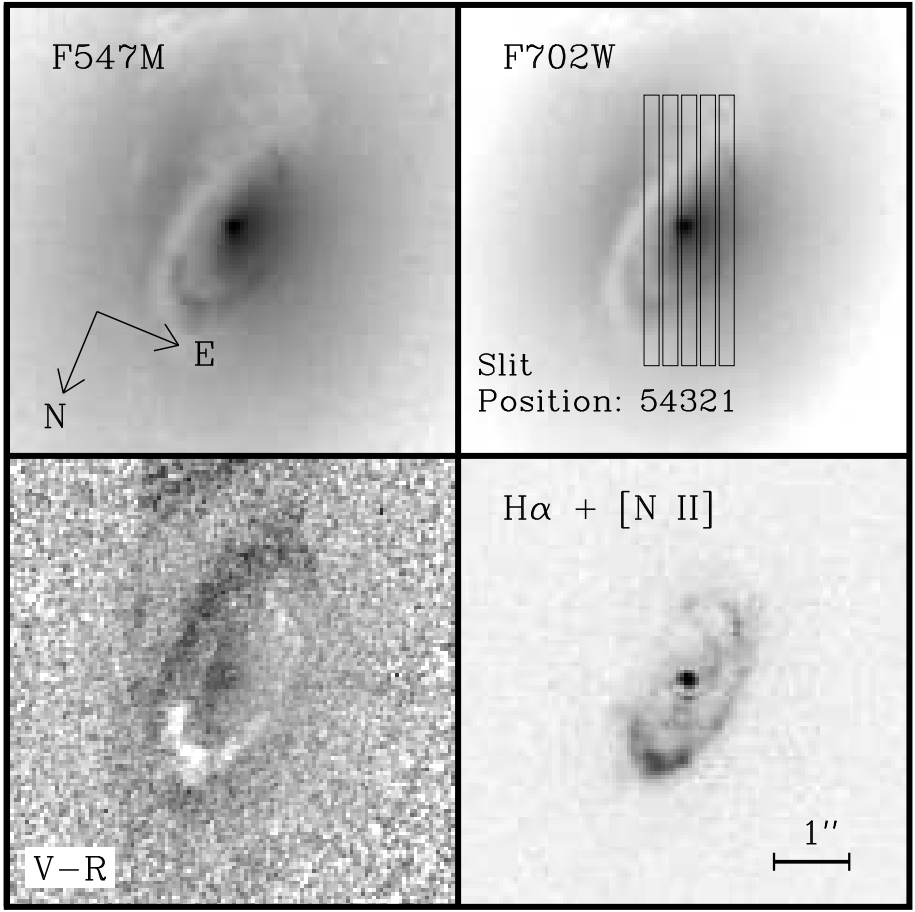}
\end{center}
\caption{\hst\ WFPC2/PC images of the nuclear region of NGC 3245.  The
images have been rotated to the STIS instrumental frame.  The
rectangles overplotted on the F702W images show the slit positions
used in the spectroscopic exposures.  Slit position 1 is the
easternmost position, and each successive spectroscopic observation
moved the slit aperture 0\farcs25 to the northwest.  The STIS
instrumental $y$ axis points upward in this figure.  In the $V-R$
image, dark regions are redder, and light regions are bluer than the
surrounding starlight.  The darkest regions correspond to a reddening
of $E(V-R) \approx 0.3$ mag relative to the unreddened regions.  Each
box is 5\farcs4 on a side. \label{figimages}}
\end{figure}

Observations of NGC 3245 were obtained with the \hst\ Wide Field and
Planetary Camera 2 (WFPC2) through the F658N (narrow-band \hn) and
F702W ($R$-band) filters, with the nucleus placed on the PC detector
(pixel size 0\farcs046).  In addition, an archival image through the
F547M ($V$-band) filter was available.  The total exposure times were
360, 2200, and 140 s for F547M, F658N, and F702W, respectively.  The
F547M and F658N images were each taken as sequences of two individual
exposures, and the frames for each filter were combined with rejection
of cosmic rays using the STSDAS task CRREJ.  Cosmic rays were removed
from the single F702W image by clipping deviant pixels using the IRAF
task COSMICRAYS.\footnote{IRAF is distributed by the National Optical
Astronomy Observatories, which are operated by the Association of
Universities for Research in Astronomy, Inc., under cooperative
agreement with the National Science Foundation.}  A
continuum-subtracted \hn\ image was created by converting the F658N
and F702W images to flux units using the \hst\ photometric zeropoints
and subtracting the scaled F702W flux from the F658N image.  The
nuclear disk is seen in dust absorption in the $V$ and $R$ images,
with an overall appearance very similar to the dusty disks seen in
some radio-loud ellipticals such as NGC 4261 \citep{jaf93} and NGC
7052 \citep{vv98}.  The disk is clearly visible in the \hn\ image as
well, with a bright, compact core of emission surrounded by an outer
ring with a major axis radius of 1\farcs1 (Figure \ref{figimages}).

From visual inspection of the images, the disk has major and minor
axes of projected size 1\farcs1 and 0\farcs4, corresponding to a
radius of 110 pc and an inclination of $\sim65\arcdeg$ to the line of
sight under the assumption that the disk is intrinsically circular.
Due to the ``fuzzy'' morphology of the disk edges, the inclination
cannot be determined precisely from the imaging data alone.  The disk
major axis and the outer isophotes of the host galaxy are aligned to
within a few degrees and are oriented approximately north-south, at
position angle (P.A.) $\approx 0\arcdeg$.

The F547M and F702W images were used to create a $V-R$ color map,
displayed in Figure \ref{figimages}.  This image shows that the dust
extinction primarily affects the western half of the disk.  Filaments
of dust are seen extending southward from the disk to distances of
$\sim3\arcsec$ from the nucleus.  The brightest arcs of \hal\ emission
at the edge of the disk coincide with regions which are bluer than the
surrounding starlight in $V-R$, which suggests that recent star
formation could be responsible for the \hn\ emission in the outer
regions of the disk.

\subsection{Stellar Density Profile}
\label{stellarprofile}

To investigate the central mass concentration in NGC 3245, it is first
necessary to determine the stellar mass density by deprojecting the
observed surface brightness distribution.  The galaxy's surface
brightness appears to be affected by dust absorption within the inner
$\sim2\arcsec$, particularly on the western side of the galaxy.  We
used the F702W data to determine the stellar light profile since it is
less sensitive to dust absorption than the F547M image.

As a first step, we removed the small contribution of emission-line
light from the F702W image, by subtracting a scaled copy of the
continuum-subtracted \hn\ image from the F702W frame, accounting for
the exposure time, photometric zeropoint, and passband difference
between the two filters.  The emission-line contribution to the total
count level in the F702W frame is highest in the ring of emission at
the outer edge of the disk.  In this region, the emission lines
contribute $3-4\%$ of the total F702W count level; in other portions
of the disk where the \hn\ lines have lower equivalent widths the
contamination is at the level of $1-2\%$.  At most locations in the
disk, however, the [\ion{N}{2}] \lam6584 line falls outside the
passband of the F658N filter.  This leaves a small residual
contamination of emission-line light in the corrected F702W image.
(This effect is somewhat mitigated by the fact that the highest
equivalent widths are found on the northern, blueshifted side of the
disk, where [\ion{N}{2}] is partially within the F658N band.)  Also,
the [\ion{S}{2}] \lam\lam6716, 6731 emission remains as a small
contribution to the F702W image.  (The [\ion{O}{1}] \lam\lam6300, 6363
lines are much fainter than [\ion{S}{2}] and we ignore their
contribution to the F702W image.)  We estimated the magnitude of this
residual contamination by extracting spectra from our STIS data and
using the F702W transmission curve to determine contribution of the
[\ion{N}{2}] and [\ion{S}{2}] lines to the F702W image at different
locations in the disk.  At most locations in the disk, the residual
emission-line contamination is $\lesssim0.7$\% of the total F702W
countrate.  Our spectra do not have the necessary spatial resolution
to perform an exact correction for this emission, but it is
sufficiently small that it has a negligible effect on the derived
stellar density profile.  In any case, the extinction within the
nuclear disk is large enough that the uncertainty in the extinction
correction (described below) is much greater than the residual
emission-line contamination.

We attempted to correct the data for the effects of dust absorption
using the technique described by \citet{car97}, using the $V-R$ color
as an indicator of dusty regions.  This method did remove patchy dust
features to some extent, but having only the $V$ and $R$ bands did not
provide a sufficient wavelength baseline for the method to be optimal.
Instead, we adopted a simpler approach to deal with the extinction.
Using the $V-R$ image, we created a map of $E(V-R)$, and from it a map
of $A_R$, under the assumption that the observed color gradient is
entirely due to dust rather than a stellar population gradient.  Then,
assuming a uniform dust screen, we used the map of $A_R$ to correct
the $R$-band image for extinction.  In the dusty patch just west of
the nucleus, the inferred value of $A_R$ is $\sim0.7-0.9$ mag.  The
dust disk is still visible in the corrected image, but is much less
optically thick than in the original frame.

\begin{figure}
\plotone{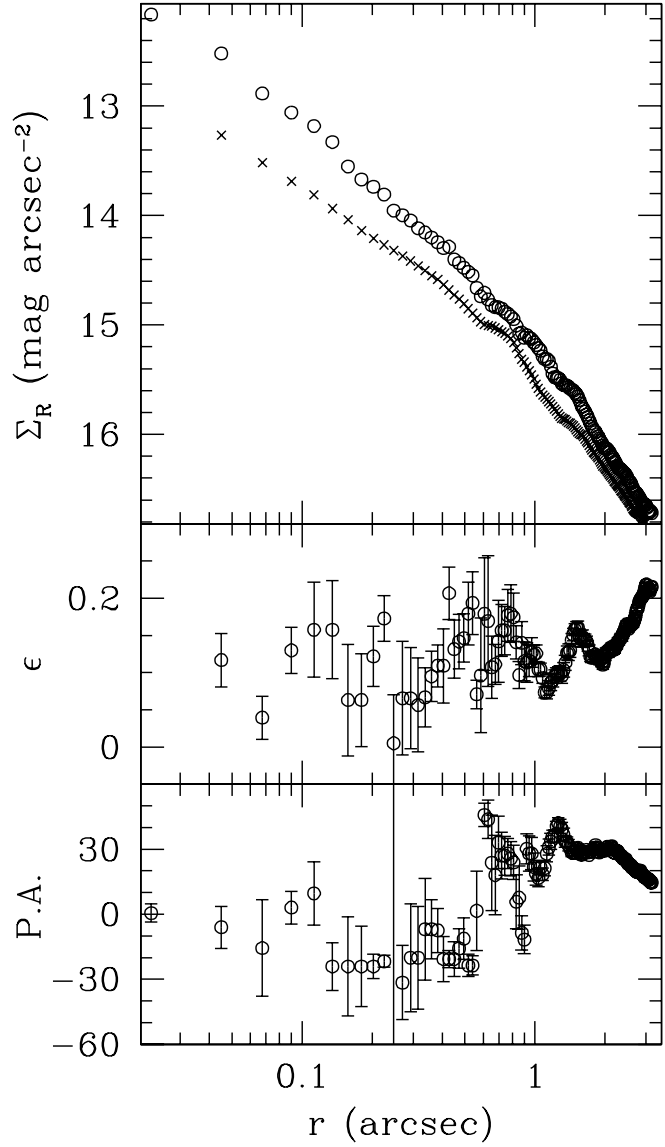}
\caption{Radial profiles of $R$-band surface brightness, ellipticity,
and position angle in NGC 3245, from the extinction-corrected WFPC2
F702W image.  The position angle is given in degrees east from north.
In the top panel, crosses represent the surface brightness prior to
the extinction correction.  Measurement uncertainties from isophote
fitting are smaller than the plot symbols in the top panel.
\label{radplot}}
\end{figure}

Surface photometry was performed on the extinction-corrected image
after masking out the remaining patches of dust located west and south
of the nucleus.  The surface-brightness profile was measured by
fitting ellipses to the isophotes of this image, using the IRAF task
ELLIPSE.  No rotations or other transformations which could
potentially blur the nuclear light distribution were performed prior
to fitting isophotes.  Within the inner $r=1\arcsec$, the isophotes of
the masked image are nearly circular, with $\epsilon \approx 0.1$
(Figure \ref{radplot}).  This enabled us to treat the surface
brightness as circularly symmetric, and the stellar distribution as
spherically symmetric, at least within the radial range where our
kinematic measurements for the ionized gas probe the galaxy potential.
 
The stellar contribution to the gravitational potential was measured
by modeling the surface-brightness profile as a sum of Gaussian
components, using the method described by \citet{mbe92} and
\citet{ems94}.  A major advantage of the multi-Gaussian method is that
it leads to a simple reconstruction of the intrinsic surface
brightness distribution, provided that the telescope point-spread
function (PSF) can be approximated as a sum of Gaussian components.
The multi-Gaussian method also leads to a straightforward deprojection
of the surface brightness $\Sigma(r)$ into the intrinsic luminosity
density distribution $\rho(r)$, which also can be expressed as the sum
of a set of Gaussians. Then, assuming a spatially constant
mass-to-light ratio $\Upsilon$, we obtain a multi-Gaussian description
for the stellar mass density as the sum of spherical mass components
having a potential $\Phi_{\star}$ that can be easily computed in terms
of error functions.  Additional examples of the method, and its
application to \hst\ images, are described by \citet{sar01}.  In the
fits, we used a multi-Gaussian description for a synthetic F702W PSF
generated by the Tiny Tim program \citep{kh99}.

The F702W count rates were converted into absolute $R$-band magnitudes
using the photometric zeropoints given by \citet{hol95} and adopting
the $V-R$ color within $r = 8\arcsec$ given by \citet{mp00}. For the
conversion between surface brightness and mass, we also corrected for
Galactic extinction using the dust map of \citet{sfd98} combined with
the reddening curve of \cite{ccm89}. The Galactic $R$-band extinction
toward NGC 3245 is $A_R = 0.067$ mag.  

We performed a decomposition of the stellar profile into two
components: the spatially extended starlight, and a central,
unresolved source.  From the decomposition, we find that the nuclear
point source has $M_R = -12.3$ mag.  We assume that this component
represents the nonstellar continuum of the weakly active nucleus.
The deprojection of the stellar profile into a three-dimensional
luminosity density was performed after subtracting the pointlike
nuclear component from the total profile.

\begin{figure}
\begin{center}
\scalebox{0.6}{\includegraphics{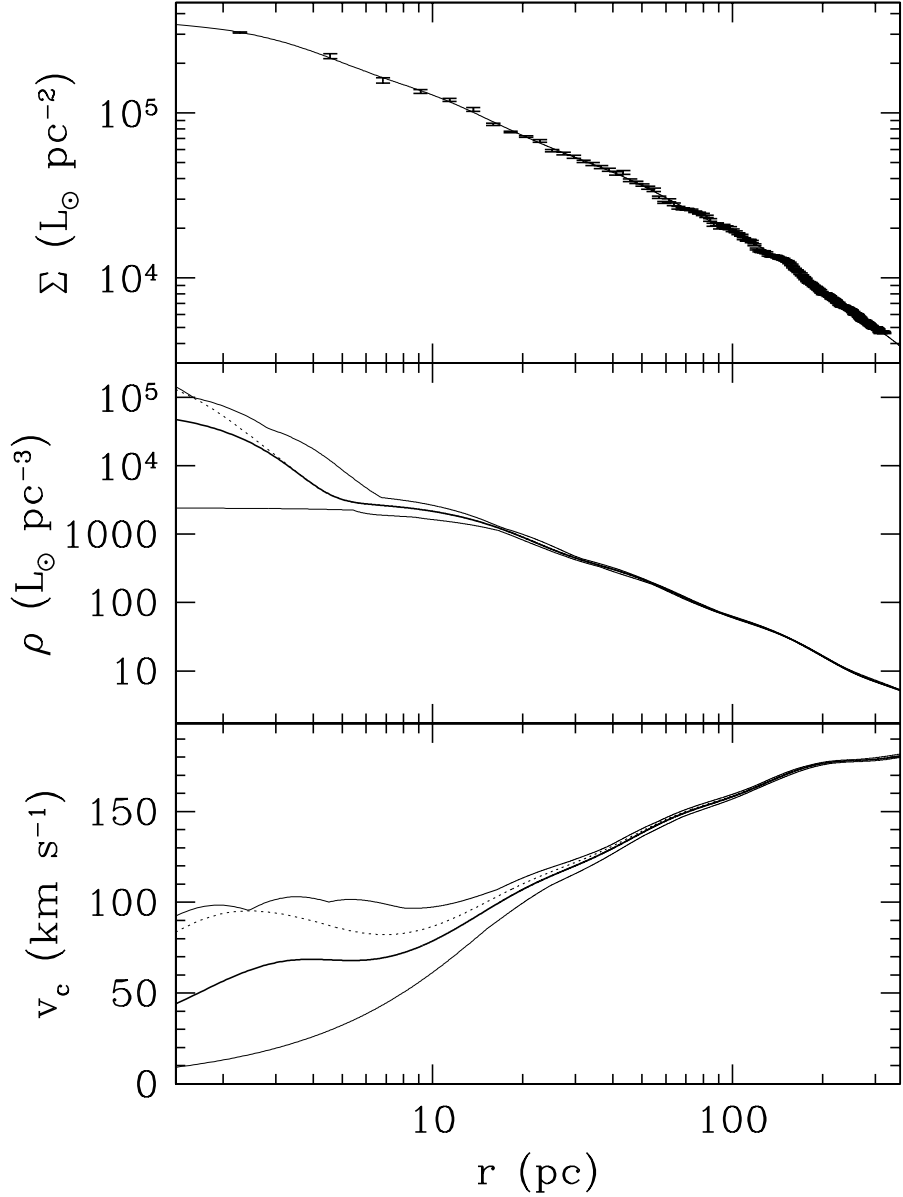}}
\end{center}
\caption{Deprojection steps for the stellar mass profile of NGC 3245.
\emph{Top panel:} Multi-Gaussian fit to the observed
surface-brightness distribution $\Sigma(r)$.  \emph{Middle panel:}
Deprojected intrinsic stellar luminosity profile $\rho(r)$ (thick
line), and $3\sigma$ confidence limits on $\rho$ (thin lines).
\emph{Bottom panel:} Circular velocity profile $v_c(r)$ obtained from
$\rho(r)$ assuming $\ml=1$ in $R$-band solar units, along with
$3\sigma$ confidence limits.  In the middle and bottom panels, the
dotted line shows the contribution of the nuclear point source, if it
is assumed to be a stellar cluster rather than nonstellar emission
from the active nucleus.
\label{deprojection}}
\end{figure}

The fit to the observed surface brightness $\Sigma(r)$, the
deprojected luminosity density profile $\rho(r)$, and the
corresponding circular velocities for $\ml=1$ (in $R$-band solar
units) are shown in Figure \ref{deprojection}.  Since the extinction
correction is derived from the ratio of the $V$ and $R$ images, the
dust-corrected image is extremely noisy, particularly in the innermost
pixels.  As a result, it is not possible to accurately measure the
cusp slope in the inner $r \approx 2$ pixels (or $\sim10$ pc).  The
deprojected stellar density profile has a broad bump in the inner 4 pc
which cannot be entirely attributed to the nuclear point source.  This
is an unavoidable problem because of the large extinction correction.
However, the uncertainty in the central density slope ultimately has a
very small effect on the dynamical models because at these small radii
the gravitational potential is dominated by the BH.

The mass of the nuclear disk itself can be estimated from the
reddening observed in the $V-R$ image.  Following \citet{ff99}, we
estimate the dust mass per unit surface area in the disk as $A_B
\Gamma_B^{-1}$, where $\Gamma_B$ is the visual mass absorption
coefficient \citep[$8\times10^{-6}$ mag kpc$^2$ \msun\per;][]{sg85}.
To estimate the maximum possible mass of the disk, we use the
reddening derived from the dusty southwestern quadrant of the disk.
For a mean reddening of $E(V-R)$ = 0.18 mag, we obtain $A_B$ = 0.93
mag and a dust mass of $1.9\times10^4$ \msun\ for the disk.  Then,
applying a Galactic gas-to-dust mass ratio of $1.3\times10^2$, the
total disk mass is $2.4\times10^6$ \msun.  This is small enough to be
safely neglected in the dynamical calculations.  The mass of the
ionized gas in the disk can be estimated from the \hal\ luminosity and
the density derived from the [\ion{S}{2}] lines
\citep[e.g.,][]{ost89}.  Using the values of $n_e$ and $L($H$\alpha)$ from
\citet{hfs97a}, we obtain a mass of $\sim5\times10^4$ \msun\ for the
ionized gas in the disk.

The mass of warm ($\gtrsim 20$ K) dust in the galaxy can be calculated
using \emph{IRAS} fluxes \citep{you89,gdj95}.  Following the method
described by \citet{you89}, we find that the total warm dust mass in
NGC 3245 is $\sim2.5\times10^5$ \msun.  This is similar to the dust
masses estimated for other early-type galaxies such as NGC 4261
\citep[e.g.,][]{mar00}.  However, the mass derived by this method
applies to the galaxy as a whole, rather than just the nuclear disk,
because of the large beamsize of the \emph{IRAS} observations.

\section{STIS Spectroscopy}
\label{stis}

\subsection{Observations}

Long-slit spectra were obtained with STIS using the 0\farcs2-wide slit
and the G750M grating, which gives a point-source spectral resolution
of FWHM $\approx0.9$ \AA.  The STIS CCD was read out in unbinned mode,
with a pixel scale of 0.554 \AA\ along the dispersion direction and
0\farcs0507 along the spatial axis in the final calibrated image.  The
grating tilt was set to give wavelength coverage over 6300--6860 \AA.
Prior to taking the spectroscopic exposures, target acquisition and
peak-up were performed to center the spectrograph slit on the nucleus.

Following the peak-up, the slit was offset by 0\farcs5 to the
southeast, and spectroscopic images were taken at five parallel slit
positions with a gap of 0\farcs05 between adjacent positions and with
the third exposure at the location of the peak-up.  The exposure times
for each position are listed in Table 1.  Figure \ref{figimages} shows
the slit locations overplotted on the WFPC2 F702W image.  The
spectrograph was oriented with the slit at P.A. = 202\arcdeg, in the
sense that the top of the slit (the instrumental $+y$ direction) was
rotated 202\arcdeg\ eastward from celestial north (see Figure
\ref{figimages}).  This orientation is offset by approximately
28\arcdeg\ from the major axis of the galactic isophotes.

Observations of internal line lamps were obtained during each orbit
for wavelength calibration.  The wavelength scale was found to shift
by 0.02--0.05 \AA\ between successive calibration frames.  The small
shifts of the wavelength scale during the course of each orbit can be
neglected, as they are much smaller than the measurement uncertainties
in the velocity centroids.

\begin{figure}
\begin{center}
\scalebox{0.9}{\includegraphics{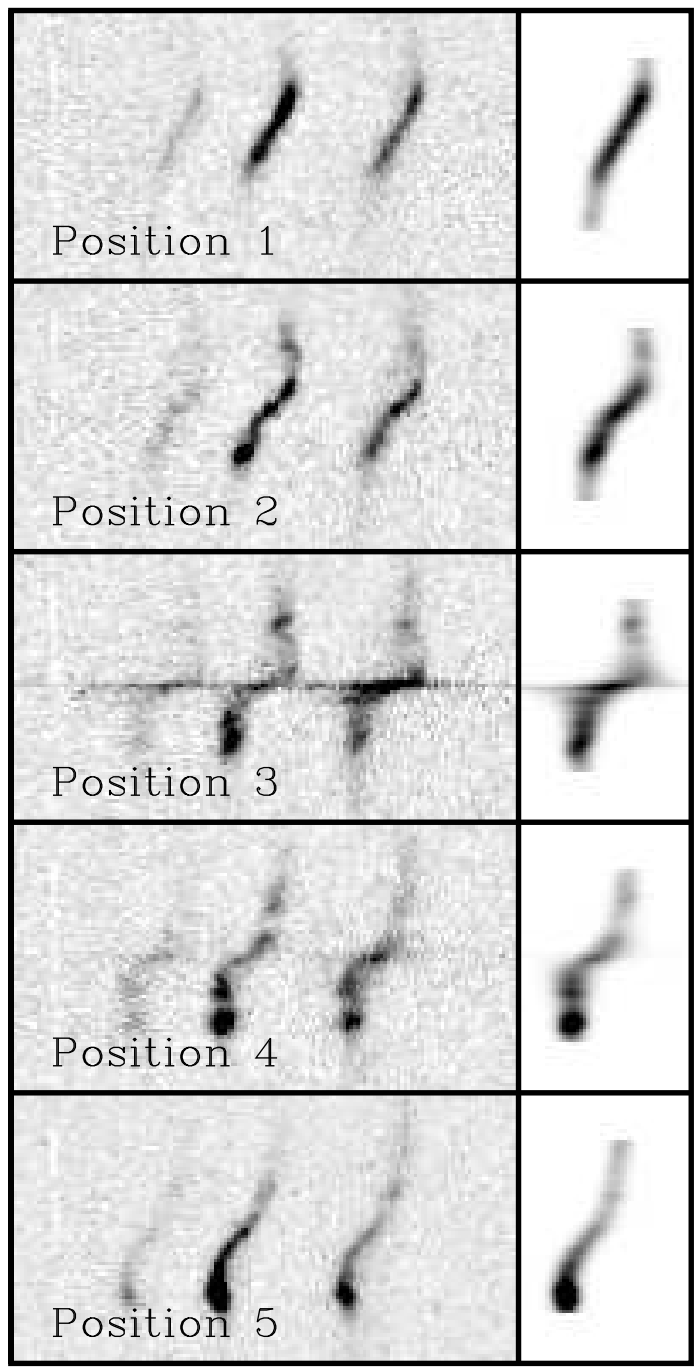}}
\end{center}
\caption{\emph{Left panels:} Portions of the geometrically rectified
two-dimensional STIS spectra showing the \hn\ emission blend.  The
spatial axis is vertical with the same orientation as in Figure
\ref{figimages}, and wavelength increases to the right.  Continuum
emission from the galaxy bulge has been subtracted.  Each box has
dimensions 4\farcs0 in the spatial direction and 83 \AA\ in the
dispersion direction.  \emph{Right panels:} The two-dimensional
synthetic spectra for each slit position for the best-fitting disk
model.
\label{figstis}}
\end{figure}

\begin{figure}
\begin{center}
\plotone{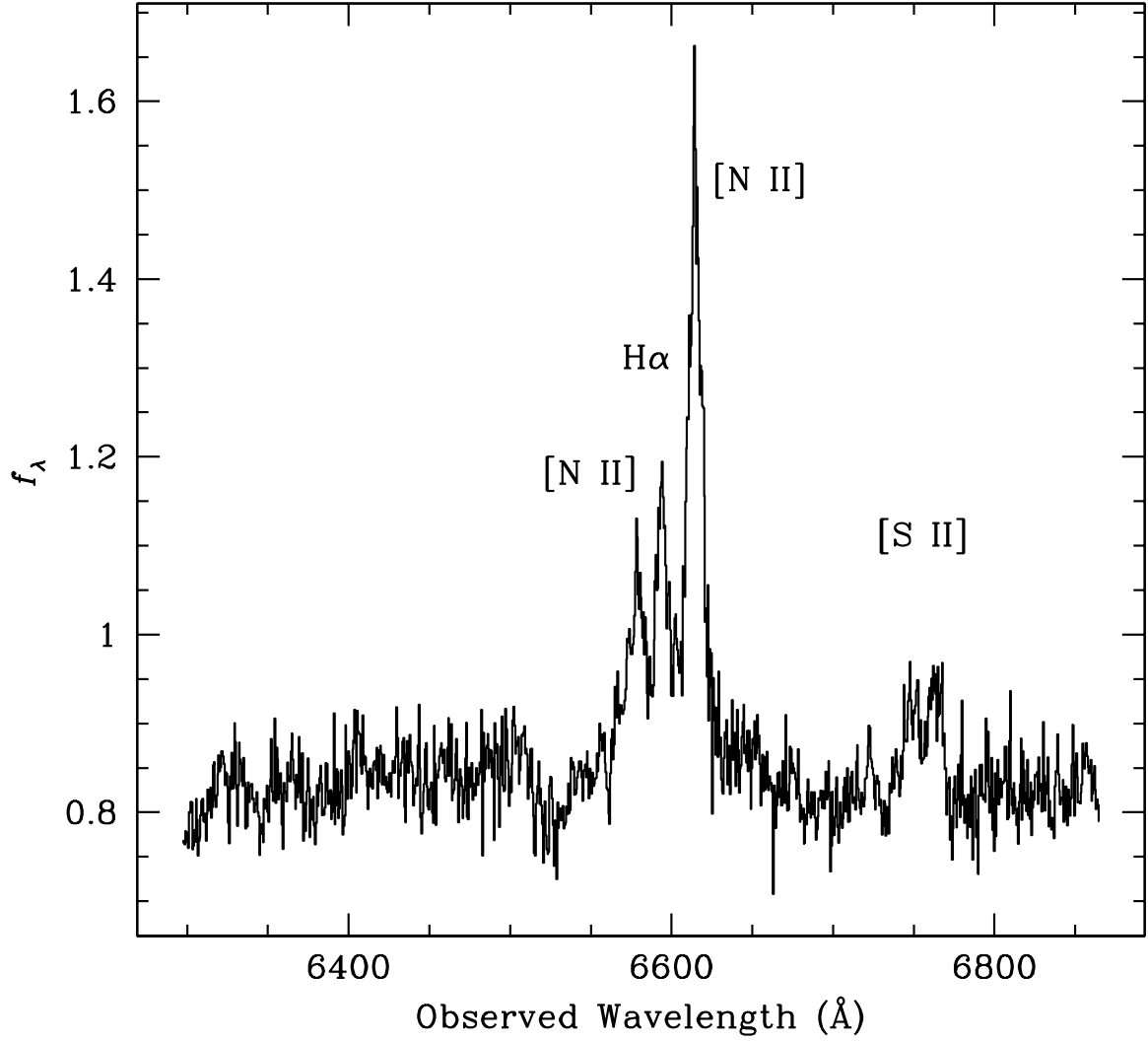}
\end{center}
\caption{The nuclear spectrum of NGC 3245, from slit position 3.  This
extraction is the sum of five CCD rows (0\farcs25 total) centered on
the nucleus, with no background subtraction applied.  The vertical
axis is in units of $10^{-16}$ erg cm\persq\ s\per\ \AA\per.
\label{nucspect}}
\end{figure}

At each slit position, three individual exposures were taken for
purposes of cosmic-ray removal.  Six total orbits were allocated for
the five slit positions.  We attempted to balance the exposure times
between the five slit positions within the constraints of a predefined
\hst\ offsetting pattern, but the use of the predefined pattern
resulted in a substantially longer exposure for the fifth slit
position.  The data were processed by the standard STIS calibration
pipeline, which includes bias subtraction, flat-fielding, cosmic-ray
removal and stacking of individual exposures, and wavelength and flux
calibration.  The pipeline calibration also corrects the wavelength
scale to the heliocentric reference frame.  The final step of the
calibration pipeline is a geometric correction that spatially
rectifies the image to a linear plate scale in the spatial and
dispersion directions.  Before applying the geometric correction, we
performed an additional step of cleaning residual hot pixels and
cosmic-ray hits from the flux-calibrated images.  The central portion
of each spectral image is displayed in Figure \ref{figstis}; the
starlight continuum has been subtracted from the data so that the
emission-line structure is clearly visible at the nucleus.  Figure
\ref{nucspect} shows an 0\farcs25-wide extraction of the nuclear
spectrum from slit position 3.

\subsection{Measurement of Emission Lines}

The emission-line velocities, linewidths, and fluxes were measured
from the STIS images by fitting a triplet of Gaussians to the \hn\
emission blend at each position.  Individual rows were extracted out
to a distance of 1\farcs5 from the slit center. At larger distances
the emission-line brightness dropped off rapidly, as expected from the
disk structure seen in the WFPC2 \hn\ image.  The profile fits were
performed independently on each single-row extraction.  Prior to
fitting, the underlying starlight was subtracted in a rudimentary way
by fitting a straight line to the continuum regions 6520--6540 \AA\
and 6660--6680 \AA, and subtracting the fitted continuum from the
spectrum.  Over this wavelength range, a straight line gives an
adequate fit to the continuum shape.

In the fits, the [\ion{N}{2}] \lam\lam6548, 6584 lines were
constrained to have the same FWHM in velocity units and a flux ratio
of 2.96:1.  All three lines were constrained to have the same mean
velocity, but the \hal\ and [\ion{N}{2}] lines were allowed to have
different widths.  For the central slit at positions within $\pm1$ row
of the nucleus, the \hn\ blend revealed a broad base of emission, and
a Gaussian broad \hal\ component was added to improve the
fits.\footnote{\citet{hfsp97b} list NGC 3245 as having an
``ambiguous'' detection of broad \hal, from their ground-based
spectra.}  Over most of the disk, the Gaussian model provided a
reasonable fit to the line profiles within the uncertainties.  At some
locations the lines were more sharply peaked than a Gaussian, but the
fits still generally resulted in \chisq\ per degree of freedom
(\chisqdof) near unity.  The lines [\ion{O}{1}] \lam\lam6300, 6363 and
[\ion{S}{2}] \lam\lam6716, 6731 are visible as well, but at much
fainter levels.  Little kinematic information could be derived from
these lines, and our analysis is based exclusively on the measurements
of \hal\ and [\ion{N}{2}].

At many locations, the redshifts of \hal\ and [\ion{N}{2}] were
discrepant by an amount larger than the formal uncertainties in the
line velocities derived from profile fits.  This is one indication
that there may be at least two kinematically distinct populations of
clouds within the disk; the two emission lines have systematically
different widths as well (\S\ref{sectionlinewidth}).  The fitting
technique described above assumes that \hal\ and [\ion{N}{2}] share
the same redshift, so the measurements essentially give a weighted
mean of the individual \hal\ and [\ion{N}{2}] line velocities as the
best estimate of the projected line-of-sight rotation velocity.

\begin{figure}
\begin{center}
\scalebox{0.7}{\includegraphics{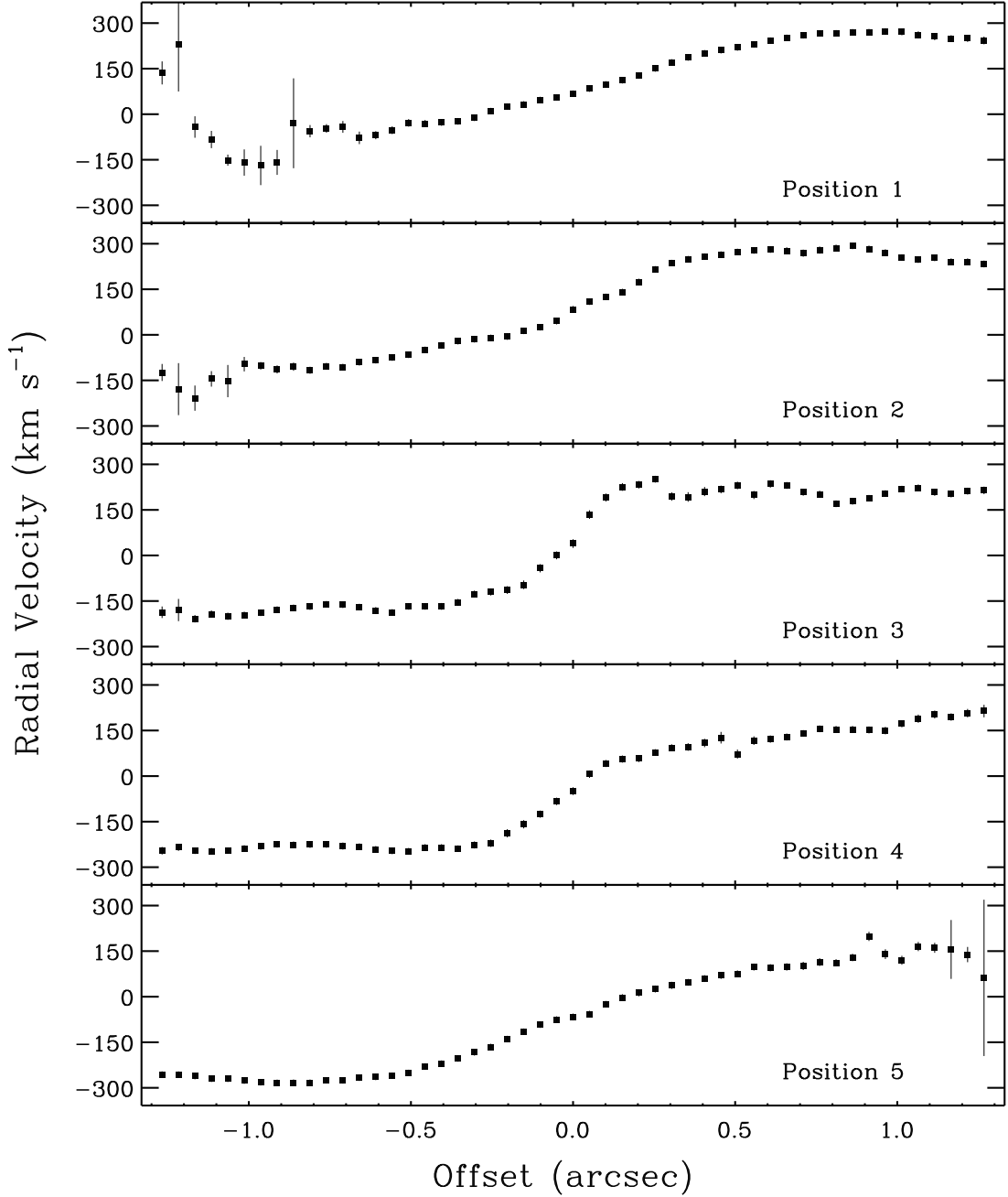}}
\end{center}
\caption{Velocity curves for the five slit positions.  The zeropoint
of the radial velocity scale is the galaxy's systemic velocity of 1388
\kms.  The left side of this plot corresponds to the bottom of the
slit (i.e., the northern side of the disk) in Figure \ref{figimages}.
\label{rotcurves}}
\end{figure}

Figure \ref{rotcurves} displays the radial velocity curves measured
from the \hn\ lines.  While there is a steep velocity gradient across
the innermost 0\farcs4 of slit position 3, we do not detect a
Keplerian rise in velocity in the innermost regions.  The steep but
smooth central velocity gradient is the result of the blurring of the
intrinsic velocity field by the telescope PSF, as well as the blending
of light from different locations in the disk within the 0\farcs2-wide
aperture.  To derive an accurate BH mass, it is necessary to model the
influence of the telescope and instrument optics (\S\ref{modeling}),
so that the data can be adequately compared with dynamical models for
the rotating disk.

\begin{figure}
\plotone{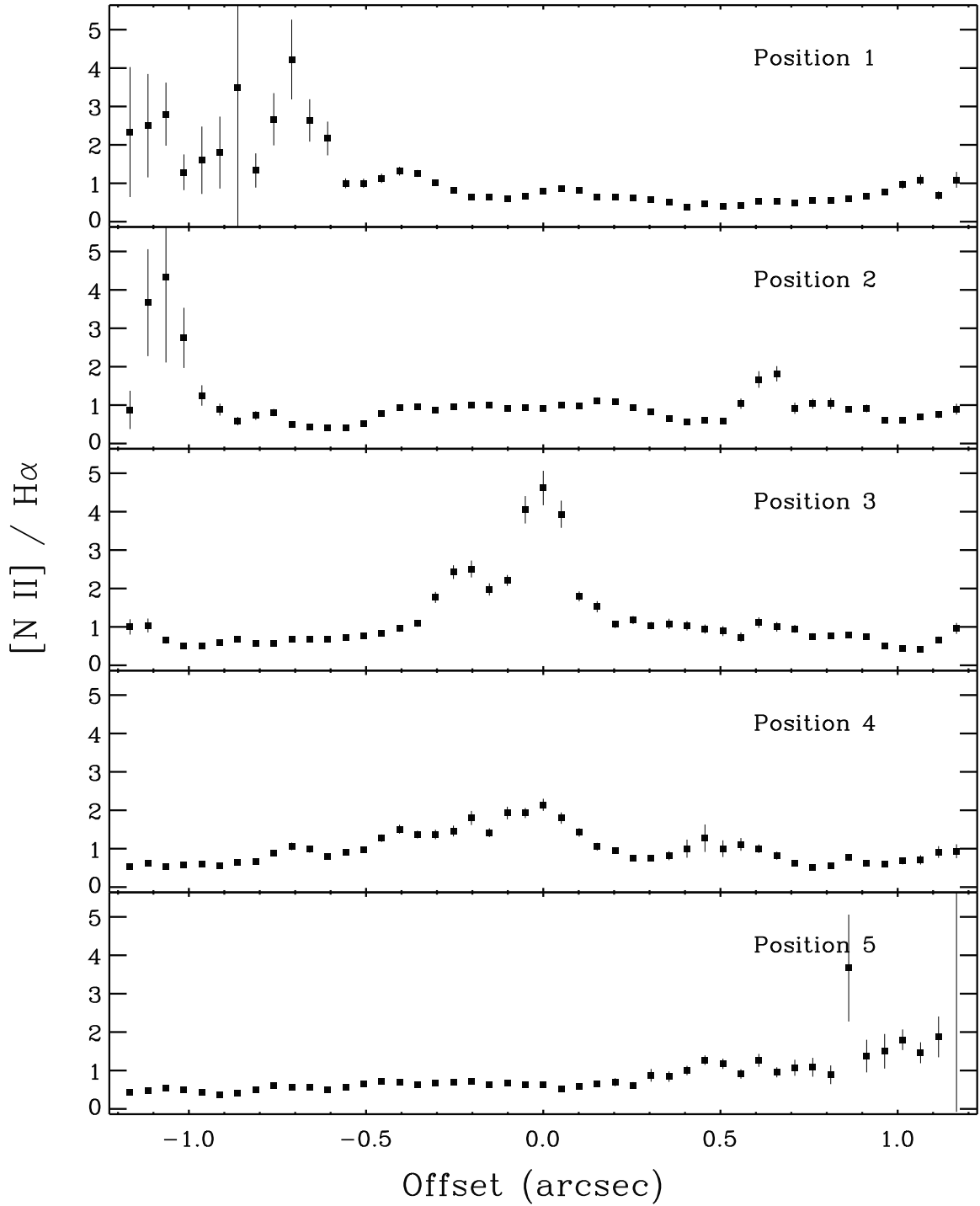}
\caption{The [\ion{N}{2}] \lam6584 / \hal\ emission-line intensity
ratio at each position in the STIS spectra.  The broad component of
\hal\ (detected only in the innermost three pixels of Position 3) is
not included.
\label{figlineratio}}
\end{figure}

Inspection of the 2-dimensional spectra reveals a gradient in the
intensity ratio of \hal\ to [\ion{N}{2}] across the disk.  For
example, Figure \ref{figstis} shows that in slit position 3, \hal\ is
more prominent than [\ion{N}{2}] in the outer disk while [\ion{N}{2}]
is stronger relative to \hal\ at the nucleus.  The [\ion{N}{2}]
\lam6584/\hal\ intensity ratio at each point is shown in Figure
\ref{figlineratio} (excluding the broad component of \hal\ at the
nucleus).  This ratio is relatively constant at $\sim0.5-0.6$ in the
outer disk but rises steeply within $r = 0\farcs3$, reaching a peak of
4.5 at the nucleus.  Such high [\ion{N}{2}]/\hal\ ratios indicate that
the ionized gas in the inner disk is excited either by photoionization
by a hard (AGN-like) spectrum \citep[e.g.,][]{hfs93}, or by shock
heating, similar to the case of the nuclear disk in M87 \citep{dop97}.
In the outer disk, some of the ionized gas may be in \ion{H}{2}
regions surrounding young, massive stars, and the region of blue
continuum emission around the outer rim of the disk further supports
this possibility.  The [\ion{O}{1}] \lam6300 line is almost
undetectable at most locations, consistent with the classification of
NGC 3245 as a ``transition''-type galaxy with properties intermediate
between LINERs and \ion{H}{2} nuclei.

\section{Velocity Field Modeling: Method}
\label{modeling}

\subsection{Basic Steps}

Modeling the observed velocity field proceeds in two steps.  First, a
model velocity field is generated and projected onto the plane of the
sky.  Assuming a thin disk in perfect circular rotation, the velocity
field is determined by the BH mass \mbh, the stellar density profile
and the stellar mass-to-light ratio \ml, and the disk inclination
\inc.  Transformation to the observed coordinate system, with
projected radii in arcseconds, depends on the distance.  Next, the
model is synthetically ``observed'' in a way that simulates as closely
as possible the actual configuration of the STIS observations.  The
synthetic observation of the model depends on the angle $\theta$ of
offset between the disk major axis and the position angle of the STIS
slit, and on the location of each slit position.  The BH mass is
determined by finding the model parameters which produce the best
match to the observed velocity curves.  This technique is similar to
the methods described by \citet{mac97}, \citet{vv98}, and \citet{vk00}
for \hst\ FOC and FOS data, and by \citet{ber98} for analysis of
ground-based data.

We use cylindrical coordinates ($r, \phi, z$) for the disk, and assume
that the disk is confined to the ($r, \phi$) plane.  The circular
velocity $v_c$ at a given radius $r$ is determined by \mbh\ and by the
total stellar mass enclosed by the circular orbit.  If the $y^\prime$
axis corresponds to the major axis of the disk on the plane of the sky
and the origin is taken to be the center of rotation, then the radius
at a projected location $(x^\prime,y^\prime)$ on the sky plane is
given by
\begin{equation}
r^2 = \frac{(x^{\prime})^2}{\cos^2 \inc} + (y^{\prime})^2.
\end{equation}
At each point $(x^\prime, y^\prime)$ in the model grid, the circular
velocity (relative to the systemic velocity \vsys) is given by
\begin{equation}
v_c = \left(\ml \vstar^2 +
\frac{G\mbh}{r}\right)^{\frac{1}{2}},
\label{eqnv}
\end{equation}
where \vstar\ is the circular velocity at radius $r$ that would result
from the stellar population alone in the absence of a BH, as
determined from the profile fits to the WFPC2 data, normalized to a
stellar mass-to-light ratio of unity.  The mass-to-light ratio \ml\
includes the contribution of spatially distributed dark matter.  We
assume that the mass-to-light ratio of the extended massive component
(stars plus dark matter) is spatially constant over the spatial scale
of the model ($r \leq 100$ pc).

In the STIS coordinate system, the $y$ axis corresponds to the
direction along the slit length and the $x$ axis is the direction
of the slit width.  The $(x^\prime,y^\prime)$ and $(x, y)$ coordinate
systems are related by the transformation
\begin{eqnarray}
x^\prime = x\cos\theta - y\sin\theta, \\
y^\prime = x\sin\theta + y\cos\theta. \nonumber
\end{eqnarray}

\subsection{Model Calculation}
\label{sectioncalculation}
The synthetic observation of the model velocity field for each STIS
CCD row is essentially a summation of the intrinsic line-of-sight
velocity profiles at every point in the disk, weighted by the
emission-line surface brightness of the disk and by the telescope PSF.
The 0\farcs2-wide slit projects to a width of almost exactly four STIS
CCD pixels.  For a given CCD row $y$, let $(x_0, y)$ be the point in
the model velocity field at the left side of the slit.  Then, the
intersection of the slit aperture and the CCD row will subtend pixels
from $x_0$ to $x_1 = x_0+3$ in the model velocity field.  We assume
that the intrinsic line-of-sight velocity profiles are Gaussian prior
to passage through the telescope optics.  Then, the line profile in
velocity units for this slit position and CCD row is given by
\begin{equation}
f_y(v) = \sum_{x=x_0}^{x_1}\sum_{i}\sum_{j} S_{ij} P(i,j|x,y)
\exp{\left({\frac{[v - v_p(i,j) - \vd M_a (x - x_c)]^2}{-2(\sigma_p^2
+ \sigmath^2 + \sigmalsf^2)}}\right)},
\label{eqnbigsum}
\end{equation}
where $S_{ij}$ is the \hn\ surface brightness at pixel $(i,j)$,
$P(i,j|x,y)$ is the value at pixel $(x, y)$ of a PSF centered at pixel
$(i,j)$, and $v_p(i,j)$ is the projected line-of-sight velocity at
$(i,j)$.  The quantity \vd\ is the bin size of the CCD pixels along
the dispersion axis in the wavelength range of interest, expressed in
velocity units ($25.2$ \kms), $x_c$ is the $x$ position of the slit
center, and ($x-x_c$) is in units of pixels.  The velocity offset $\vd
M_a (x - x_c)$ is the shift due to the nonzero width of the slit and
its projection onto the STIS CCD; it accounts for the fact that the
wavelength recorded for a photon depends on the position at which the
photon enters the slit along the $x$ axis.  The anamorphic
magnification factor $M_a$ gives the ratio of plate scales in the
dispersion and spatial directions.  In the wavelength range of
interest, $M_a = 0.93$ for the G750M grating \citep{bb98}.  The
quantity $\sigma_p$ is a possible bulk velocity dispersion for gas in
the circumnuclear disk, projected along the line of sight, and
\sigmath\ is the thermal velocity dispersion in the disk.  The \hn\
surface brightness profile $S$ is discussed in \S\ref{sectionsurfbr},
and the derivations of $v_p$ and $\sigma_p$ are discussed in
\S\ref{sectionlinewidth} and \S\ref{sectionadrift}.  The quantity
$\sigmalsf$ is the Gaussian width of the point-source line-spread
function (LSF) for STIS (\S\ref{sectionlinewidth}).

The summations over $i$ and $j$ are assumed to extend over the entire
model velocity field, as the extent of the PSF is comparable to the
angular size of the NGC 3245 disk.  We found, however, that omitting
the extended PSF wings, beyond a radius of about 0\farcs4, had a
negligible effect on the model results.  Since the computation time
increases steeply with increasing PSF size, we performed our model
calculations using a subsection of the full PSF with dimensions
$0\farcs9\times0\farcs9$.  The omitted flux from the extended PSF
wings amounts to only 5\% of the total flux in the PSF.

\begin{figure}
\begin{center}
\includegraphics{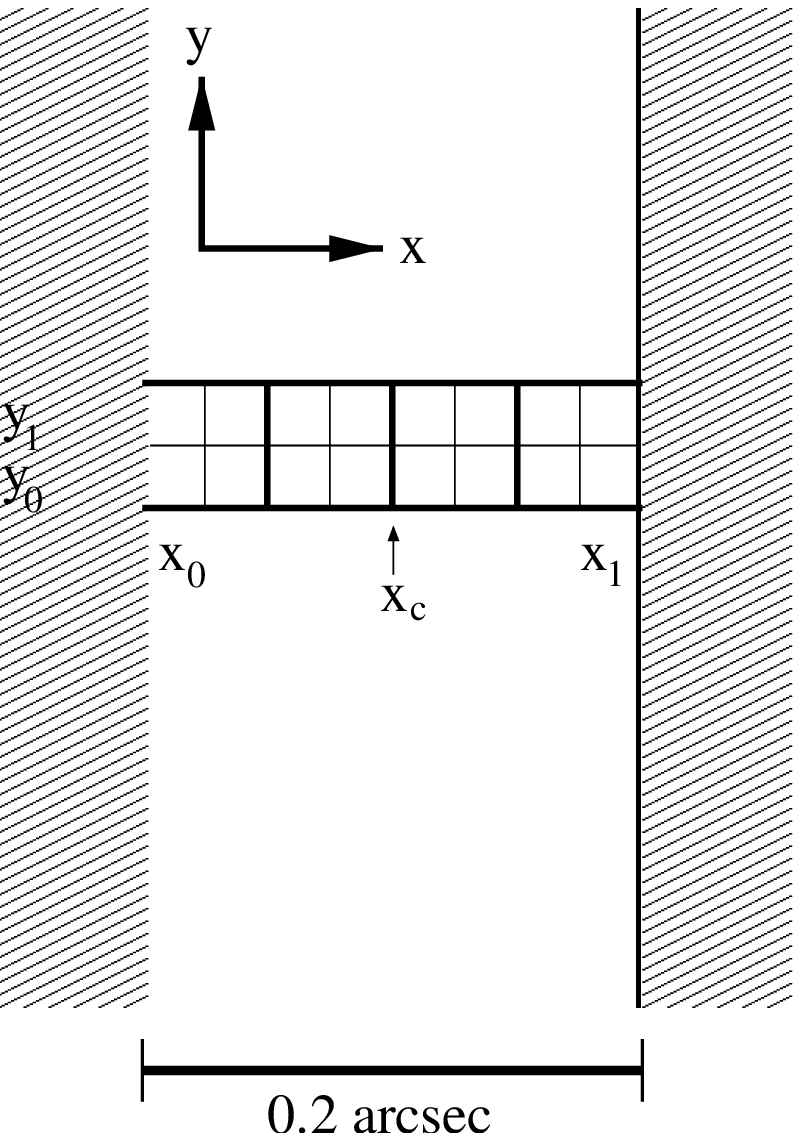}
\end{center}
\caption{Diagram illustrating the slit geometry used in the
line-profile calculations (equation \ref{eqnbigsum}).  For a given
STIS CCD row, the predicted line profile is computed by summing the
contributions of light entering the slit through a $4s \times s$ grid
of subpixel elements, where $s$ is the subsampling factor.  The light
entering each subpixel element includes contributions from each point
in the model velocity field, weighted by the \hn\ surface brightness
and the PSF.  The slit midpoint in the $x$ direction is located at
$x_c$. The diagram shows the case $s=2$.
\label{slitdiagram}}
\end{figure}

In practice, the entire procedure described above was calculated on a
subsampled pixel grid, with a subsampling factor $s$ ranging from 2 to
6 relative to the STIS pixel scale.  Thus, each model subpixel element
was $0\farcs0507/s$ on a side.  Especially at locations near the
nucleus, where there are large gradients both in velocity and in \hal\
surface brightness, a higher value of $s$ can in principle yield a
more accurate model calculation.  On a subsampled pixel grid,
calculating the line profile for a given CCD row requires a summation
over $s$ rows of subpixel elements (i.e., over subsampled rows $y_0$
to $y_1 = y_0 + s - 1$).  This adds one additional level of summation
to equation (\ref{eqnbigsum}); see Figure \ref{slitdiagram} for an
illustration.  In the limit of high values of $s$, this summation
converges to an integration over infinitesimal pixels, similar to the
expressions given in equation (10) of \citet{mac97} or equation (1) of
\citet{ber98}.  Since the PSF and the \hn\ surface brightness are
defined on a discrete pixel grid, the actual calculation that we
perform is a summation over pixels rather than an integration of
analytic functions.

We used Tiny Tim PSFs for STIS, which were generated on subsampled
pixel grids to match the resolution of the model calculations.  The
PSFs were computed for a monochromatic source at 6600 \AA.  Two
distinct physical effects contribute to the PSF: an optical component
resulting from diffraction of light through the \hst\ and STIS
apertures, and a charge-diffusion component which arises from the
bleeding of electrons from each CCD pixel into adjacent pixels.  These
two components must be treated separately in the modeling process,
because the optical PSF will scatter light into the slit from every
position within the PSF radius as in equation (\ref{eqnbigsum}) above,
while the charge-diffusion component of the PSF will only cause a
spreading of charge from each pixel into its immediate neighbors.  The
charge-diffusion portion of the PSF accounts for a significant
fraction of the total PSF: roughly 20\% of the charge incident on a
given pixel will bleed into the adjoining pixels \citep{kh99}.  Thus,
the quantity $P$ in equation (\ref{eqnbigsum}) denotes only the
optical component of the PSF and not the charge-diffusion portion,
which is applied at a later stage in the model calculation.  With the
Tiny Tim program, PSFs generated on subsampled pixel grids contain
only the optical component by default.

PSFs determined from STIS imaging observations could in principle be
used as well, but would suffer from several disadvantages: limited
signal-to-noise ratio (S/N) and dynamic range; a lack of information
on the sub-pixel structure of the PSF for subsampled models; and the
inability to disentangle the optical and charge-bleeding components.
In addition, STIS has no narrow-band \hal\ filter, so any PSF
determined from imaging observations would be obtained with either a
clear aperture or a long-pass filter, and its structure would differ
significantly from the monochromatic PSF appropriate for the
spectroscopic data.  One drawback of Tiny Tim PSFs for STIS is that
they do not include spatial variations in focus aberrations for
different positions on the CCD \citep{kh99}. However, in the NGC 3245
STIS data, \hal\ falls almost exactly at the center of the CCD, where
the Tiny Tim PSF should be most accurate.

The synthetic line profiles were calculated on a velocity grid with a
bin size of 25.2 \kms\ in order to match the STIS pixel scale in the
relevant wavelength range.  The coarseness of this wavelength scale
has no effect on the derived velocity centroids or the linewidths of
the synthetic profiles (to better than 1 \kms) because the model
profiles are noise-free.  For each slit position, the array of model
line profiles forms a synthetic 2-dimensional spectrum, analogous to
the STIS spectroscopic data and with the same pixel scale (Figure
\ref{figstis}).  The 2-dimensional model profiles were then convolved
with the CCD charge-diffusion kernel to simulate the bleeding of
charge between adjacent pixels.  This kernel is approximated by a
$3\times3$ pixel array given by \citet{kh99} for the STIS CCD.  This
convolution only results in very small changes (a few \kms) to the
radial velocity curves and linewidths, but it is included for
completeness.

Finally, to simulate the measurement of the actual STIS data, each
synthetic line profile was measured by extracting individual rows and
fitting a Gaussian model, analogous to the measurement of the actual
STIS data.  The synthetic profiles were nearly Gaussian in shape
except within the inner $\sim0\farcs3$, where the large velocity
gradient at the nucleus produced asymmetries and extended
low-amplitude wings on the profiles.  In principle, it would be
simpler and computationally faster to calculate a weighted average
velocity for the light entering the slit \citep[for example,
see][]{mac97}, rather than computing a line profile and performing a
profile fit.  However, we found that for slit positions near the
nucleus, the weighted average velocity gave misleading results because
of the asymmetric wings on the line profiles.  Fitting the calculated
profiles with a Gaussian ensures that the data and models are treated
identically.  Also, computing the full line profile allows a
comparison with the observed line widths, in addition to the velocity
curves.

One aspect of the data which we do not model is the geometric
rectification of the image.  This correction spatially resamples the
image to rebin it to a linear pixel scale (using bilinear
interpolation), and rectifies the image so that the dispersion
direction is aligned parallel to the CCD rows.  As a result, the
spatial and wavelength scales of the corrected image are expanded in
some locations and compressed at other locations, relative to the
original image.  Our model calculation is performed on a spatially
rectified grid with a linear pixel scale, and there is no simple way
to account for the spatially variable effects of the geometric
correction.  To check whether the geometric correction affects the
linewidths, we measured the width of [\ion{N}{2}] from the uncorrected
image.  There are small differences in the linewidths measured from
the corrected and uncorrected images, but overall the geometrically
corrected image does not show systematically broader lines, indicating
that the geometric correction should not have a significant effect on
the derived value of \mbh.

\subsection{Emission-Line Surface Brightness}
\label{sectionsurfbr}

\begin{figure}
\scalebox{0.7}{\includegraphics{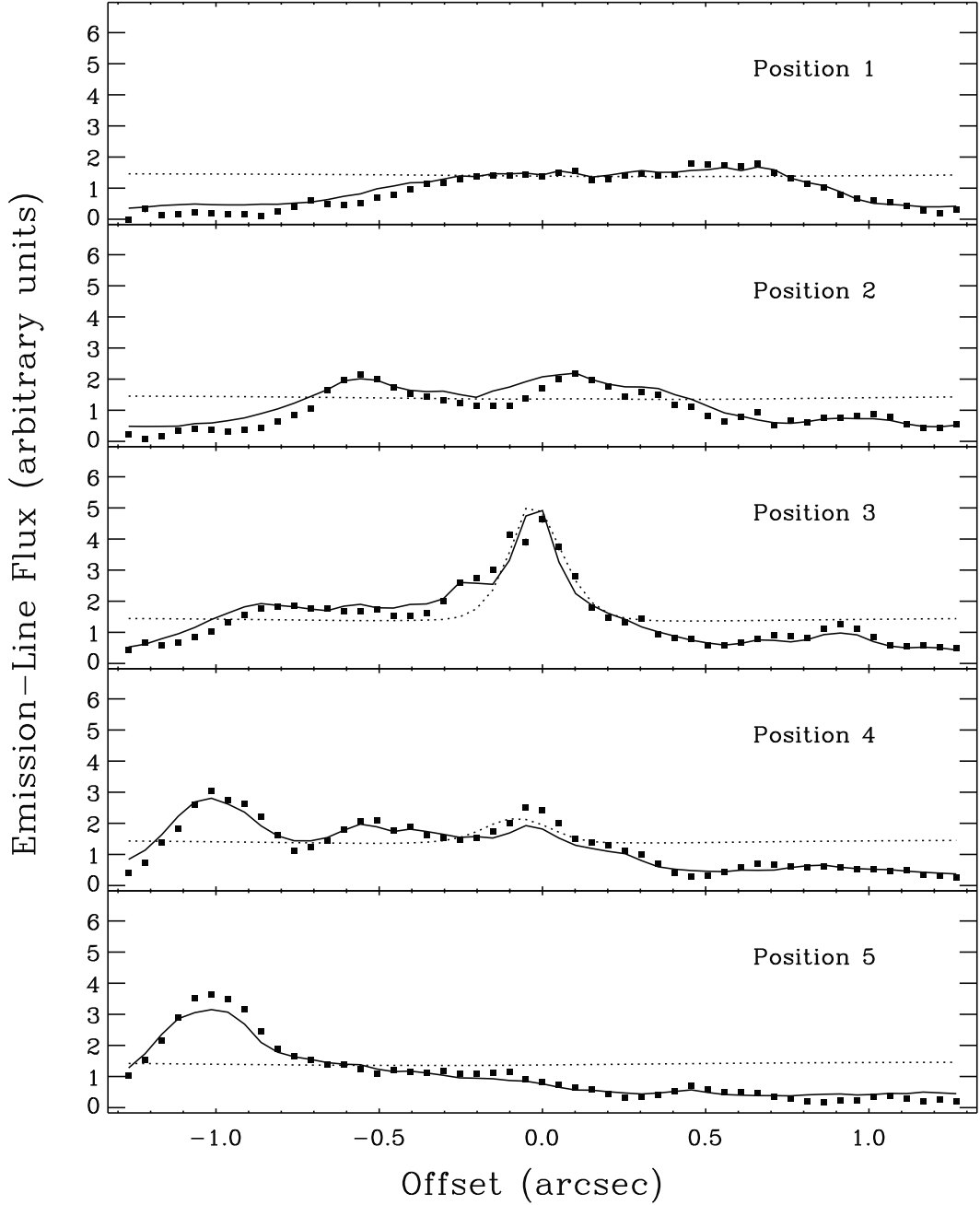}}
\caption{The total \hn\ emission-line flux measured at each position
in the STIS spectra.  \emph{Solid curves:} Model results using the
continuum-subtracted, deconvolved WFPC2 F658N image as the \hn\
surface brightness.  A single scaling factor has been applied in order
to match the median values of the STIS data and the model
curves. \emph{Dotted curves:} Results of calculations based on the
exponential surface-brightness model. \label{figflux}}
\end{figure}

\begin{figure}
\rotatebox{-90}{\scalebox{0.65}{\includegraphics{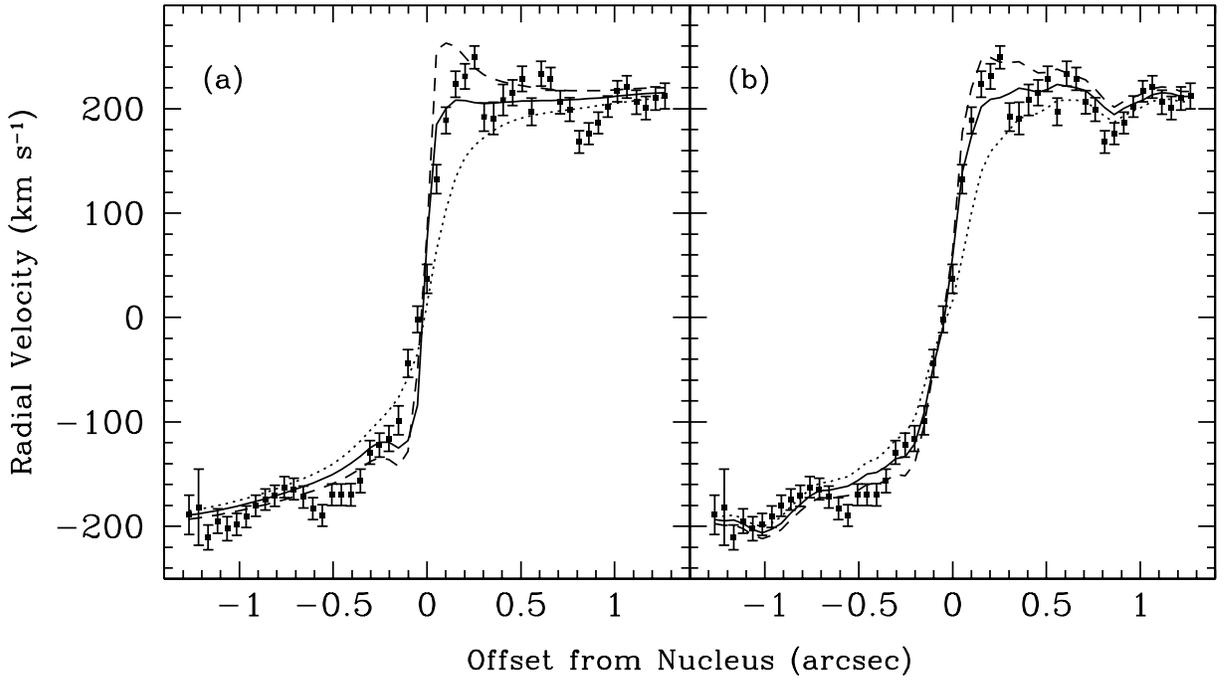}}}
\caption{Radial velocity curves for slit position 3.  (a) Models
computed using the exponential model to approximate the \hn\ surface
brightness.  (b) The same models, but computed using the \hn\ surface
brightness measured from the continuum-subtracted, deconvolved WFPC2
F658N image.  BH masses are 0 (\emph{dotted line}), $2\times10^8
\msun$ (\emph{solid curve}), and $4\times10^8 \msun$ (\emph{dashed
curve}).  The models are computed for \ml\ = 3.5 \mlsun\ to illustrate
the fit to the outer portions of the disk, but the inner $r
= 0\farcs5$ would be better fit with \ml\ in the range 3.6--3.8
\mlsun.
\label{figmiddle}}
\end{figure}

In the model calculation, the line profiles must be weighted by the
emission-line surface brightness at each point in the disk.  One
option is to use a smooth analytic approximation to the observed
surface brightness, similar to the method used in most previous work
\citep[e.g.,][]{vv98, vk00}.  As an initial step, we parameterize the
\hn\ surface brightness by $S(r) = S_0 + S_1 e^{-r/r_1}$.  (We will
refer to this as the exponential surface-brightness model.)  The three
parameters $S_0$, $S_1$, and $r_1$ were determined by fitting a model
to the inner $r = 0\farcs5$ of the disk and optimizing the model fit
to the spatial profile of \hn\ flux in the STIS data.  The results of
the fit are shown in Figure \ref{figflux}.  We find $S_0 = 1$ and $S_1
= 26$ (in arbitrary units), and $r_1 = 6.3$ pc; similar results are
found by fitting directly to the surface brightness profile of the
WFPC2 \hn\ image.  Folding the exponential surface-brightness profile
into the model calculation results in smooth velocity curves, as shown
in the left panel of Figure \ref{figmiddle}.

For NGC 3245, the WFPC2 \hn\ image provides sufficient S/N and spatial
coverage that the image itself can be folded into the model
calculations, rather than using a smooth model for the surface
brightness.  Even though [\ion{N}{2}] \lam6584 is mostly outside the
F658N filter band, the WFPC2 image can still provide an approximate
description of the \hn\ flux distribution, and it allows the models to
incorporate the small-scale structure in the emission-line surface
brightness profile.  Before performing model calculations with the
\hn\ image, we first removed the contribution of the broad \hal\
component from the nucleus, because the radial velocity curves were
measured only for the narrow components.  In the central row
extraction of slit position 3, the broad component of \hal\ accounts
for 46\% of the total emission-line flux that would be measured
through the F658N filter at this position.  This broad-line
contribution was subtracted off from the F658N image by scaling a Tiny
Tim PSF image to the appropriate flux and subtracting it from the
continuum-subtracted \hn\ image.

The emission-line map must represent the surface brightness prior to
convolution with the telescope optics, since it will effectively be
convolved with the STIS PSF in the model calculation.  To prepare the
\hn\ map, we first resampled the WFPC2 image onto a pixel grid
subsampled by a factor of \subsamp\ with respect to the original PC
pixel size (0\farcs046), and then deconvolved it using the
Richardson-Lucy algorithm using the IRAF task LUCY.  The deconvolution
was performed using a synthetic Tiny Tim WFPC2 PSF generated on a
pixel grid subsampled by a factor of \subsamp.  Beyond about 10
iterations, the deconvolution tended to amplify noise peaks to
unacceptable levels, and we halted the deconvolution at 10 iterations.

After deconvolution, the image was rotated to match the orientation of
the STIS observations, and resampled again to a pixel size which
matched the scale of the velocity field map ($0\farcs0507/s$
pixel\per).  The location of the dynamical center was assumed to be
coincident with the nuclear peak in the \hn\ image.  The pixel values
of this deconvolved image were then used for the surface brightness in
the model calculations, as described by equation (\ref{eqnbigsum}).
This method gives a good fit to the \hn\ line fluxes measured from the
STIS data over the entire face of the disk (Figure \ref{figflux}).  We
do not attempt to fit the surface-brightness distributions of \hal\
and [\ion{N}{2}] separately, as we do not have sufficient information
to disentangle the individual contributions of \hal\ and [\ion{N}{2}]
to the WFPC2 image.

By folding the observed surface brightness into the calculations, the
models are able to reproduce some of the apparent irregularities in
the observed velocity curves.  In part, this is due to the velocity
gradient in the portion of the disk subtended by the slit.  An equally
important effect is the instrumental shift in apparent wavelength for
light entering the slit off-center.  With the 0\farcs2-wide slit,
light entering at the edge of the slit will suffer an instrumental
offset of roughly 2 pixels in the $x$ direction, or $\sim50$ \kms,
compared to light entering at the slit center.  The observed velocity
curves show irregular bumps and wiggles at roughly this level, similar
to the irregularities seen in the velocity structure of other nuclear
disks \citep[e.g.,][]{sar01}.

Figure \ref{figmiddle} demonstrates that many (but not all) of the
apparent irregularities in the velocity curves are the result of the
patchy distribution of emission-line light, rather than being real
deviations from circular rotation.  This gives additional confidence
that the nuclear disk can be modeled successfully in terms of a thin
disk in circular rotation.  It also allows the identification of
localized regions which genuinely deviate from the overall circular
velocity pattern.  

One additional detail is the treatment of the central subpixel element
in the model calculation.  The rotation velocity for this central
pixel is undefined, and it is located at the peak of the emission-line
surface-brightness distribution.  We tried two methods to handle this
singular point, by setting either the rotation velocity or the
emission-line surface brightness of the central pixel to zero.  We
obtained a better fit to the data by setting the rotation velocity to
zero in the central pixel, but the resulting radial-velocity curves
were nearly identical in either case.  Since our measurements are
relatively insensitive to the very high-velocity wings on the line
profiles, the treatment of the innermost pixel makes little difference
to the model predictions as long as the model is calculated on a
sufficiently fine grid.

\subsection{Location of STIS Apertures}
\label{sectionlocation}

In the modeling, an issue of paramount importance is the determination
of the actual positions of the STIS apertures relative to the nucleus
of NGC 3245.  An error of even half a pixel in the central slit
location would have a substantial effect on the model results.  (The
radial velocity curves for the outer positions are less sensitive to
the precise location of the slit.) The observing program was nominally
designed to obtain five parallel slit positions with a separation of
0\farcs25 between adjacent pointings, and with the third pointing
centered at the nucleus.  However, the asymmetry in the radial
velocity curve for slit position 3 demonstrates that the center of
rotation must have been located off-center in the slit.

One constraint on the slit locations comes from the peak-up image,
which is an undispersed image of the galaxy as seen through the STIS
slit. The peak-up procedure is designed to maximize the amount of
light entering the slit, and the asymmetric distribution of dust at
the nucleus resulted in the galaxy's nucleus being offset with respect
to the slit center.  This miscentering is seen clearly in the peak-up
image, and from visual inspection of this image we estimate that the
brightest point at the nucleus was located $\sim1.5-2$ pixels from the
slit center after the peak-up.  

We determined the slit positions quantitatively by two methods,
comparing the continuum and emission-line light profiles of the
spectroscopic data with the imaging observations.  The continuum light
profiles in the spectra were compared with profiles measured from the
STIS acquisition image, a 5 s exposure taken with a long-pass filter
admitting light in the range 5500--10000 \AA.  For each of the STIS
spectroscopic images, a continuum light profile was calculated by
collapsing the spatial profile along the wavelength axis over a
wavelength range of 6400-6500 \AA.  Comparison profiles were created
from the acquisition image by extracting 4-pixel-wide strips oriented
along the slit direction.  The spatial position of each spectroscopic
slit was determined by computing a \chisq\ parameter for the
difference between the spectroscopic continuum profile and each
comparison profile extracted from the image.  This method confirms
that the central slit position was miscentered with respect to the
brightest pixel at the nucleus, with the best results obtained for an
offset between 1 and 2 STIS pixels between the slit $x$ center and the
nuclear peak.  The derived positions of the other slits are all
consistent with the nominal offsets of 0\farcs25 perpendicular to the
slit length between successive exposures.  There is no evidence for
any translation of the telescope along the slit direction; the maximum
possible shift between exposures is 0\farcs01--0\farcs02, or less than
0.5 STIS pixels.  

To determine the slit positions with sub-pixel accuracy, we ran a
model fit comparing the emission-line flux curves from the STIS
spectra (Figure \ref{figflux}) with the model results based on folding
the WFPC2 emission-line surface-brightness profile into the model
calculation.  The separation between adjacent slits was fixed at the
nominal value of 0\farcs05 but the zeropoint of the five slit
positions was allowed to float freely in the $x$ and $y$ directions,
and the model was calculated with $s=4$.  The model fit was optimized
by calculating a \chisq\ parameter for the difference between the
measured and modeled emission-line flux curves, as shown in Figure
\ref{figflux}.  The fit converged on a position for the galaxy nucleus
which was 1.5 pixels offset from the slit center in the $x$ direction,
in good agreement with the constraints on the slit locations from the
continuum profiles and the peak-up image.

Based on these results, we fixed the position of the central slit in
the modeling analysis so that the dynamical center of the galaxy in
slit position 3 was located 1.5 CCD pixels (0\farcs076) away from the
slit center, as measured along the instrumental $-x$ direction (the
direction perpendicular to the slit length).  In the $y$ direction,
the galaxy's dynamical center was fixed at the center of a STIS pixel,
with an uncertainty of $\sim0.25$ pixel based on the model fits to the
emission-line flux distribution.

\subsection{Linewidths}
\label{sectionlinewidth}

In this section we discuss the various effects that contribute to the
observed linewidths.  There are three major contributions:
instrumental broadening, the blending of light from different parts of
the disk within the slit aperture (i.e., rotational broadening), and a
possible intrinsic velocity dispersion in the disk itself.  For the
thermal contribution to the line broadening, we use $\sigmath = 10$
\kms, as expected for $T\approx10^4$ K.  This contribution is assumed
to be isotropic.

\begin{figure}
\begin{center}
\scalebox{0.65}{\includegraphics{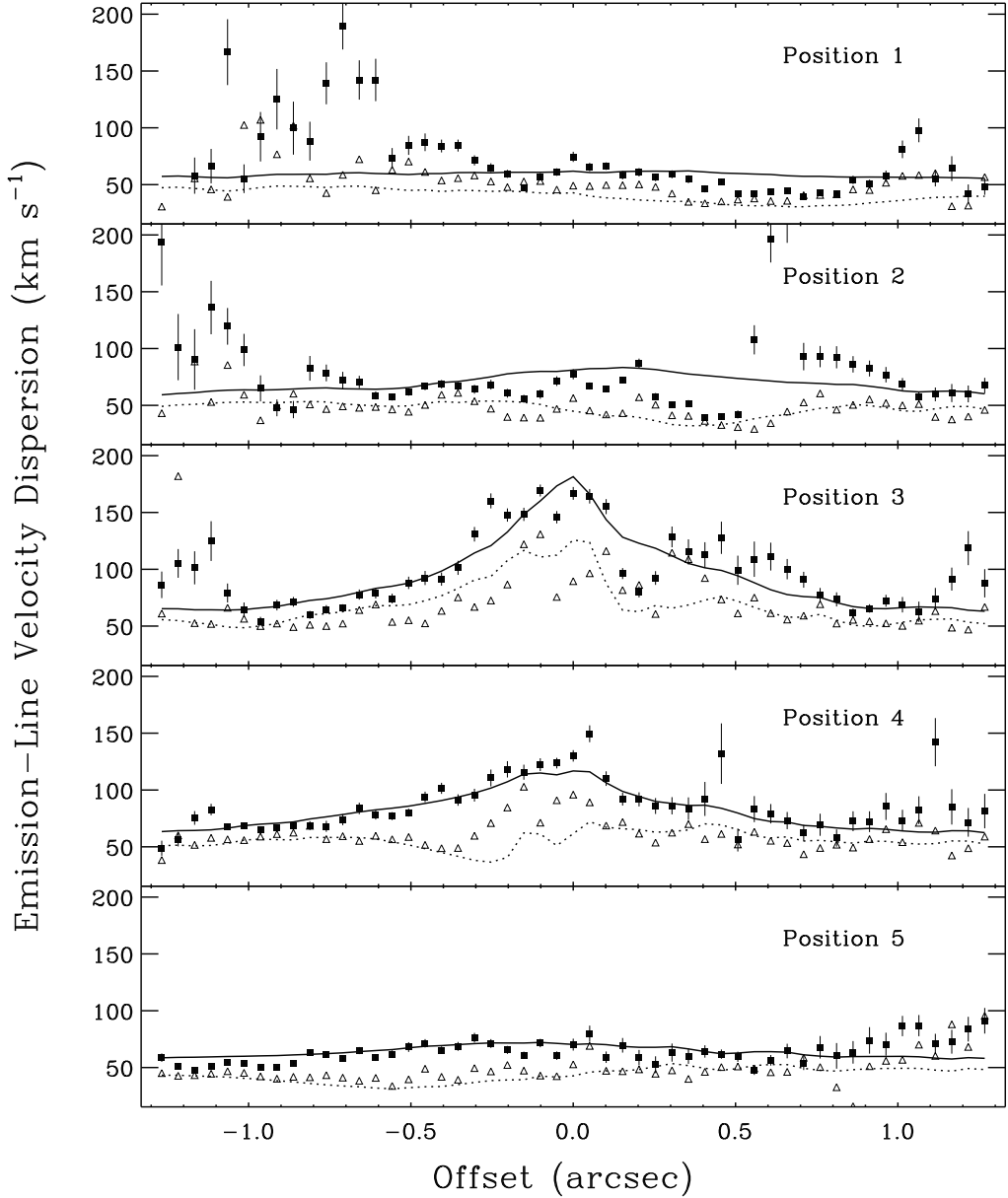}}
\end{center}
\caption{Velocity dispersions of [\ion{N}{2}] (\emph{filled squares})
and \hal\ (\emph{open triangles}) at each position in the STIS
spectra.  For clarity, error bars on the \hal\ velocity dispersions
have been omitted.  \emph{Dotted curves:} Results of model
calculations for a dynamically cold disk with zero intrinsic
linewidth.  In these models, the only contributions to the predicted
linewidths are rotational and instrumental broadening.  \emph{Solid
curves:} Model results incorporating the intrinsic velocity dispersion
in the disk, with parameters set to fit the [\ion{N}{2}] data.
Regions outside the nucleus showing very high velocity dispersions (at
the left side of positions 1 and 2) are the result of low S/N at these
positions; since velocity dispersion is always positive, random errors
at very low S/N will bias the measurements toward high values of
$\sigma$.
\label{figlinewidth}}
\end{figure}

The measured velocity dispersion of [\ion{N}{2}] (uncorrected for
instrumental broadening) increases steadily from $\sim50$ \kms\ in the
outer disk to $\sim160$ \kms\ at the nucleus.  \hal\ is narrower than
[\ion{N}{2}] throughout the disk (Figure \ref{figlinewidth}).  The
difference between the \hal\ and [\ion{N}{2}] linewidths may be partly
due to Balmer absorption in the continuum underlying \hal.  The
highest \hal\ equivalent widths (up to 30 \AA) are found at the
northern edge of the disk, and \hal\ is still systematically narrower
than [\ion{N}{2}] in this region, but the blue continuum in this
region may indicate a young stellar population which could contribute
additional Balmer absorption.  Thus, it is unclear whether \hal\ and
[\ion{N}{2}] have intrinsically different widths.  Unfortunately,
[\ion{O}{1}] and [\ion{S}{2}] are too weak to be useful for a
systematic comparison, as done for example by \citet{cor99}.  Since
\hal\ may be affected by underlying absorption, we use the
[\ion{N}{2}] linewidths as the main comparison for our model
calculations.

The instrumental LSF for a point source (including broadening by the
telescope PSF and charge diffusion on the CCD) can be approximated by
a Gaussian with $\sigma \approx 17$ \kms\ for the G750M grating at
$\lambda = 6590$ \AA\ \citep{stismanual}.  The observed line widths
are much broader than this, so it is not necessary to model the
detailed structure of the LSF.  In the model calculation, the
contributions to line broadening due to the nonzero slit width, PSF
broadening, and charge diffusion are included explicitly.  Thus, the
quantity \sigmalsf\ in Equation (\ref{eqnbigsum}) represents only the
remaining portion of the point-source LSF, which is largely due to the
roughness of the grating surface.  This quantity cannot be measured
directly but can be estimated by comparison with the linewidths
measured from wavelength calibration lamp exposures.

The STIS Instrument Handbook \citep{stismanual} lists an expected
extended-source LSF of FWHM $\approx4$ pixels for the 0\farcs2-wide
slit.  However, the comparison lamp exposures obtained with our data
showed linewidths of FWHM = 3.3 pixels.  To estimate \sigmalsf, we
computed a model for a monochromatic source of uniform surface
brightness.  The observed linewidths for the comparison lamp exposures
were reproduced by setting $\sigmalsf$ in the range 5--8 \kms.  This
represents a very small contribution to the total line broadening, so
the model results are insensitive to the exact value of \sigmalsf.

Preliminary model runs (shown in Figure \ref{figlinewidth}) indicated
that the observed [\ion{N}{2}] linewidths are too large to be merely
the result of rotational, instrumental, and thermal broadening.  In
other words, the gas in the disk must have a substantial bulk velocity
dispersion, similar to what has been observed in circumnuclear disks
in some other early-type galaxies \citep{vv98, vk00}. To match the run
of linewidths over the face of the disk, we parameterize the radial
component of this bulk velocity dispersion with the function
\begin{equation}
\sigma_r(r) = \sigma_0 + \sigma_1 e^{-r / r_0}.
\end{equation}
To find the optimal values for the three parameters $\sigma_0$,
$\sigma_1$, and $r_0$, we calculated a set of models to minimize a
goodness-of-fit parameter $\chi_\sigma^2$ that described the squared
deviation between the model and observed linewidths.  The best-fitting
values are $\sigma_0=35$ \kms, $\sigma_1=105$ \kms, and $r_0=50$ pc.
Since the intrinsic linewidth is the dominant source of line
broadening in the data, these three parameters can be fixed at these
values in the later model optimizations that we use to determine \mbh\
and \ml.  In other words, models with different BH masses and hence
different amounts of rotational broadening will still have essentially
the same final linewidths because the intrinsic linewidth is
significantly larger than the rotational broadening at most locations,
and the different sources of broadening effectively add in
quadrature.  

The \hal\ velocity dispersion, on the other hand, is reproduced
extremely well over most of the disk by a model for a cold disk with
no bulk velocity dispersion (Figure \ref{figlinewidth}).  Only at the
centers of slit positions 3 and 4 does the \hal\ velocity dispersion
rise significantly above the cold disk model predictions.  There is a
great deal of scatter in the \hal\ linewidths at the center of slit
position 3, but the central region of slit position 4 can be fit
reasonably well by a model having the same values of $\sigma_1$ and
$r_1$ as in the [\ion{N}{2}] model, but having $\sigma_0=0$.
Differences in the widths of different emission lines have been
observed in almost every ionized gas disk studied with \hst\ data.  In
some cases it has been attributed to underlying differences in the
spatial distribution of different ionic species, as a result of
spatially unresolved density and/or ionization gradients \citep[see
discussion in][]{vk00}.  However, the STIS data shows that in NGC
3245, [\ion{N}{2}] and \hal\ have different widths over the entire
face of the disk.  If this difference is not entirely due to the
effects of underlying \hal\ absorption, then there must be spatially
coincident but kinematically distinct components within the disk.

In the models for the dynamically cold disk (the dotted line in Figure
\ref{figlinewidth}), the line widths are determined only by the
instrumental, rotational, and thermal broadening.  At some positions
in the model, the gradients in rotational velocity and in instrumental
wavelength shifts across the slit are oppositely directed and
partially cancel each other out, and the predicted line profiles
become narrower than in surrounding areas.  These regions are seen as
the broad dips in velocity dispersion for the cold disk model, for
example at $-0\farcs5$ from the slit center in slit position 5.  In an
extreme case, the instrumental and rotational broadening can cancel
out almost exactly, yielding an extremely narrow linewidth in a
localized region.  \citet{mb01} discuss these features in STIS
spectra, which they refer to as ``caustics,'' and they show that the
position of a caustic can have diagnostic power for measuring the
central mass.  However, a substantial intrinsic velocity dispersion in
the gas will blur out the caustic and make it difficult to detect in
practice.  Our model calculated for a dynamically cold disk predicts
one caustic feature, at a location $-0\farcs35$ from the slit center
in slit position 4.  At this small distance from the nucleus, both
[\ion{N}{2}] and \hal\ have a large intrinsic velocity dispersion, and
the caustic is not observed in either line.

\subsection{Asymmetric Drift}
\label{sectionadrift}

In common with some other galaxies having circumnuclear disks, NGC
3245 shares the trait that the intrinsic velocity dispersion of the
emission-line gas rises dramatically toward the nucleus, at least for
[\ion{N}{2}].  This intrinsic linewidth is much larger than the
thermal width for a cloud at $T \approx 10^4$ K, so it may represent
either local turbulence within the disk, or the combined motion of a
system of collisionless clouds on eccentric orbits, similar to the
stellar-dynamical case.  It is important to address the possible
dynamical effect of this velocity dispersion in the models. If it
contributes a pressure which supports the disk against gravity, then
the observed mean rotation speed $v_\phi(r)$ will deviate from the
local circular velocity and models without an asymmetric drift
correction will underestimate the true central mass.

The origin of the intrinsic velocity dispersion in nuclear gas disks
is presently not understood, and it is not clear whether asymmetric
drift plays a role in the dynamical structure of these objects.  For
IC 1459, where the nuclear disk becomes rounder at small radii and the
innermost velocities appear to depart from a Keplerian curve,
\citet{vk00} suggest that asymmetric drift may be important.
Similarly, \citet{crd00} describe an application of an asymmetric
drift correction to the gas velocities in the elliptical galaxy NGC
2320, in order to derive the true circular velocity from the observed
rotation velocity of the gas.  On the other hand, most previous
gas-dynamical BH measurements have assumed that the gas velocity
dispersion results from a local ``microturbulence,'' and that the gas
still rotates at the circular velocity.  Van der Marel \& van den
Bosch (1998) argue that the gas velocity dispersion in NGC 7052 is due
to local turbulence and does not affect the orbital structure.
Without a clear understanding of the physical nature of the intrinsic
velocity dispersion, we choose to calculate models both with and
without an asymmetric drift correction, so that we can estimate the
resultant uncertainty in the derived value of \mbh.  The asymmetric
drift is calculated using only the [\ion{N}{2}] velocity dispersions,
rather than \hal, so as to estimate the maximum possible effect on the
mass determination.
 
We assume that motions in the gas disk are close to isotropic in the
$r$ and $z$ coordinates. Then, $\sigma_z = \sigma_r$ and
$\left<v_rv_z\right> = 0$, and the asymmetric drift correction can be
expressed as
\begin{equation}
v_c^2 - v_\phi^2 = \sigma_r^2 \left[ -r \frac{d\ln
\nu}{dr} - r\frac{d\ln\sigma_r^2}{dr} - \left(
1 - \frac{\sigma_\phi^2}{\sigma_r^2} \right) \right],
\end{equation}
where $\sigma_\phi$ is the azimuthal velocity dispersion, and $\nu(r)$
is the number density of gas clouds in the disk \citep[e.g.,][]{bt87,
gkv90}.  Since the disk surface brightness depends on proximity to the
active nucleus and/or regions of star formation, the luminosity
density in the disk is probably not a good tracer of the underlying
number density of clouds.  Nevertheless, for lack of a better
alternative, we use the emission-line surface brightness distribution
$S(r)$ to represent the cloud density.  

\begin{figure}
\begin{center}
\plotone{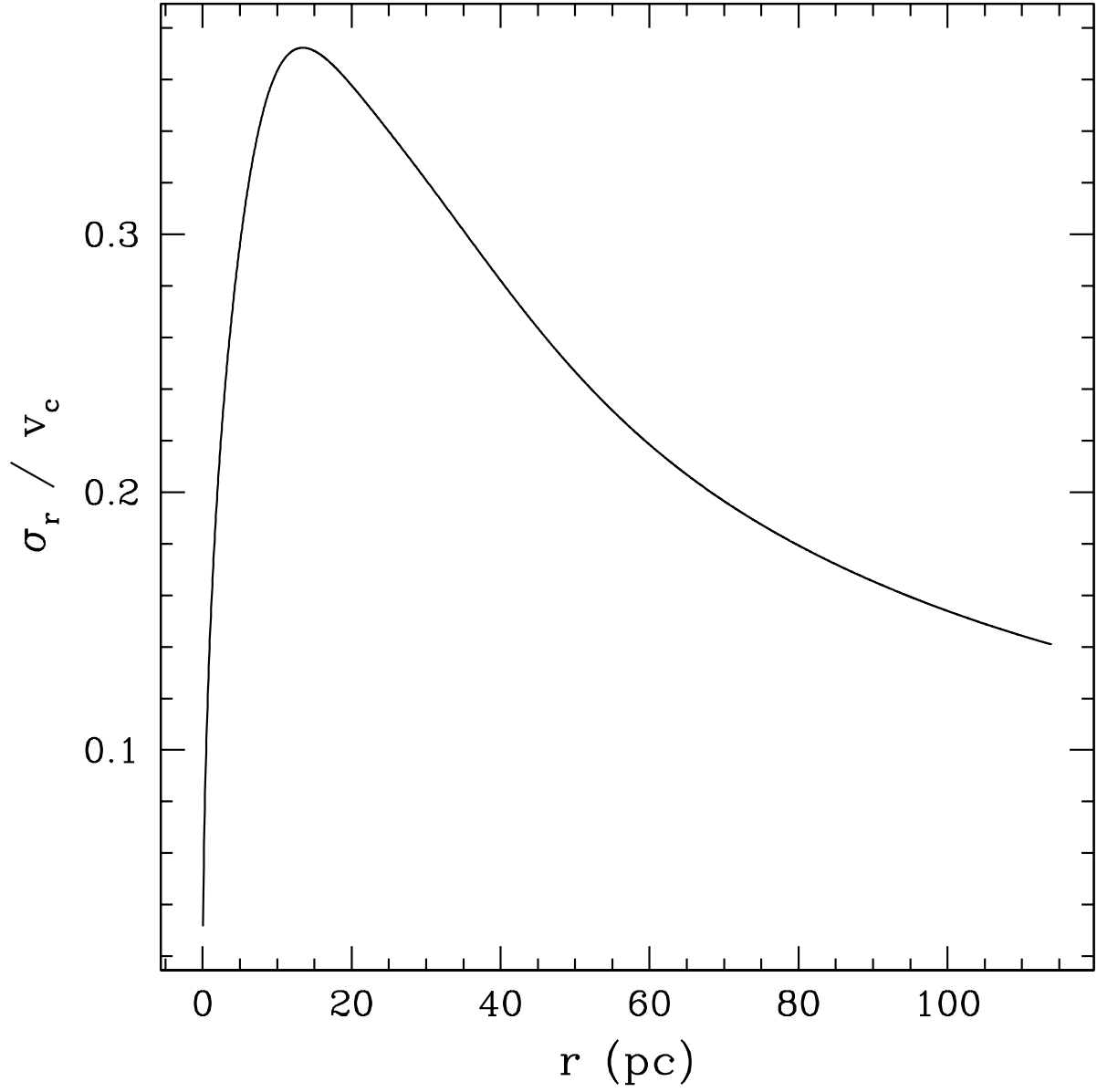}
\end{center}
\caption{The ratio $\sigma_r / v_c$ for the best-fitting model to the
NGC 3245 data.
\label{sigmaoverv}}
\end{figure}

In order to take the radial derivative of an analytic model rather
than the noisy surface brightness profile itself, we use the
exponential surface-brightness model to represent the number density
of clouds in the disk.  Using an analytic model of this kind is also
preferable because the underlying number density of clouds in the disk
is likely to be much more symmetric and more smoothly distributed than
the patchy emission-line surface brightness. The relation between
$\sigma_r$ and $\sigma_\phi$ is given by the epicycle approximation
\begin{equation}
\frac{\sigma_\phi^2}{\sigma_r^2} = \frac{-B}{A-B},
\end{equation}
where $A$ and $B$ are the Oort constants \citep{bt87}.  This
approximation is strictly valid only in the limit $\sigma \ll v_c$,
and for some other galaxies this method cannot be applied because
$\sigma \approx v_c$ in the nuclear regions \citep[e.g., IC
1459;][]{vk00}.  However, our best-fitting model for NGC 3245 has
$\sigma_r / v_c < 0.35$ everywhere in the disk (Figure
\ref{sigmaoverv}), so this method can be adopted for at least an
approximate treatment of the asymmetric drift.  A more detailed
assessment of the magnitude of the asymmetric drift would require
dynamical modeling of the orbits of gas clouds in the disk, and such
calculations are beyond the scope of this paper.  In reality, the
stellar-dynamical analogy must break down at some level because gas
clouds in the disk will suffer collisions at a rate that depends on
the typical cloud size and the filling factor of clouds in the disk.
Thus, the actual mean rotation velocity in the disk will most likely
be bracketed by the models with and without the asymmetric drift
correction.

After calculating the asymmetric drift, we obtain the mean rotation
velocity $v_\phi(r)$.  The projected line-of-sight velocity field is
calculated according to
\begin{equation}
\label{eqnvp}
v_p = v_\phi \left(\frac{y^\prime \sin\inc}{r}\right),
\end{equation}
and the projected velocity dispersion along the line of sight is
\begin{equation}
\label{eqnsigma}
\sigma_p^2 = \left( \sigma_\phi^2 - \sigma_r^2 \right) \left(
\frac{y^\prime \sin i}{r} \right)^2 + \sigma_r^2.
\end{equation}
These values of $v_p$ and $\sigma_p$ are then used in the model
calculation (equation \ref{eqnbigsum}).  For models without an
asymmetric drift correction, these relations reduce to $\sigma_p =
\sigma_\phi = \sigma_r$ and $v_\phi = v_c$, corresponding to the case
where the intrinsic velocity dispersion represents a locally isotropic
microturbulence within the disk.

\subsection{Uncertainties and \chisq\ Calculation}

To optimize the models, we calculated a \chisq\ parameter for the
model fit to the velocity curves.  The linewidths and emission-line
fluxes at each point carry little or no information about the possible
presence of a BH, so we do not include them in the \chisq\
calculation.  The \chisq\ parameter was calculated over a rectangular
region, and we define the ``fitting radius'' \rfit\ as the maximum
distance from the slit center over which \chisq\ was calculated for
each slit. 

The initial uncertainties in the measured velocities are the values
returned by the triple-Gaussian fit to the \hal\ and [\ion{N}{2}]
emission lines.  By forcing \hal\ and [\ion{N}{2}] to have the same
velocity in the fit, we effectively determine a best-fitting mean
velocity for the blend, at the expense of underestimating the true
uncertainty in the velocity measurements.  In regions of high S/N, the
formal model-fitting uncertainties are only a few \kms.  At some
locations, however, the velocities of \hal\ and [\ion{N}{2}] disagree
by up to $\sim30$ \kms.  In addition, there are regions in the disk
which systematically deviate from the circular rotation models for any
reasonable model parameters.  Since the formal error bars are much
smaller than these local deviations from pure circular rotation, no
model fit can be expected to yield $\chisqdof \approx 1$.  When
deriving a goodness-of-fit parameter to optimize the model fits, we
choose to rescale the error bars so as to incorporate these local
deviations from pure circular rotation as effectively random errors in
velocity.  This ``extra'' uncertainty is independent of the S/N of the
data, as it depends only on the dynamical properties of the disk at a
given location.  Thus, we rescale the error bars by adding a single
constant in quadrature to the velocity uncertainties.  This constant
was set to 10 \kms\ so that the best-fitting model with \rfit\ =
0\farcs5 would yield $\chisqdof\ \approx 1.0$.  Rescaling the error
bars in this fashion is conservative in the sense that it increases
the final uncertainty and widens the confidence intervals on \mbh.

\section{Velocity Field Modeling: Results}
\label{results}

\begin{figure}
\begin{center}
\plotone{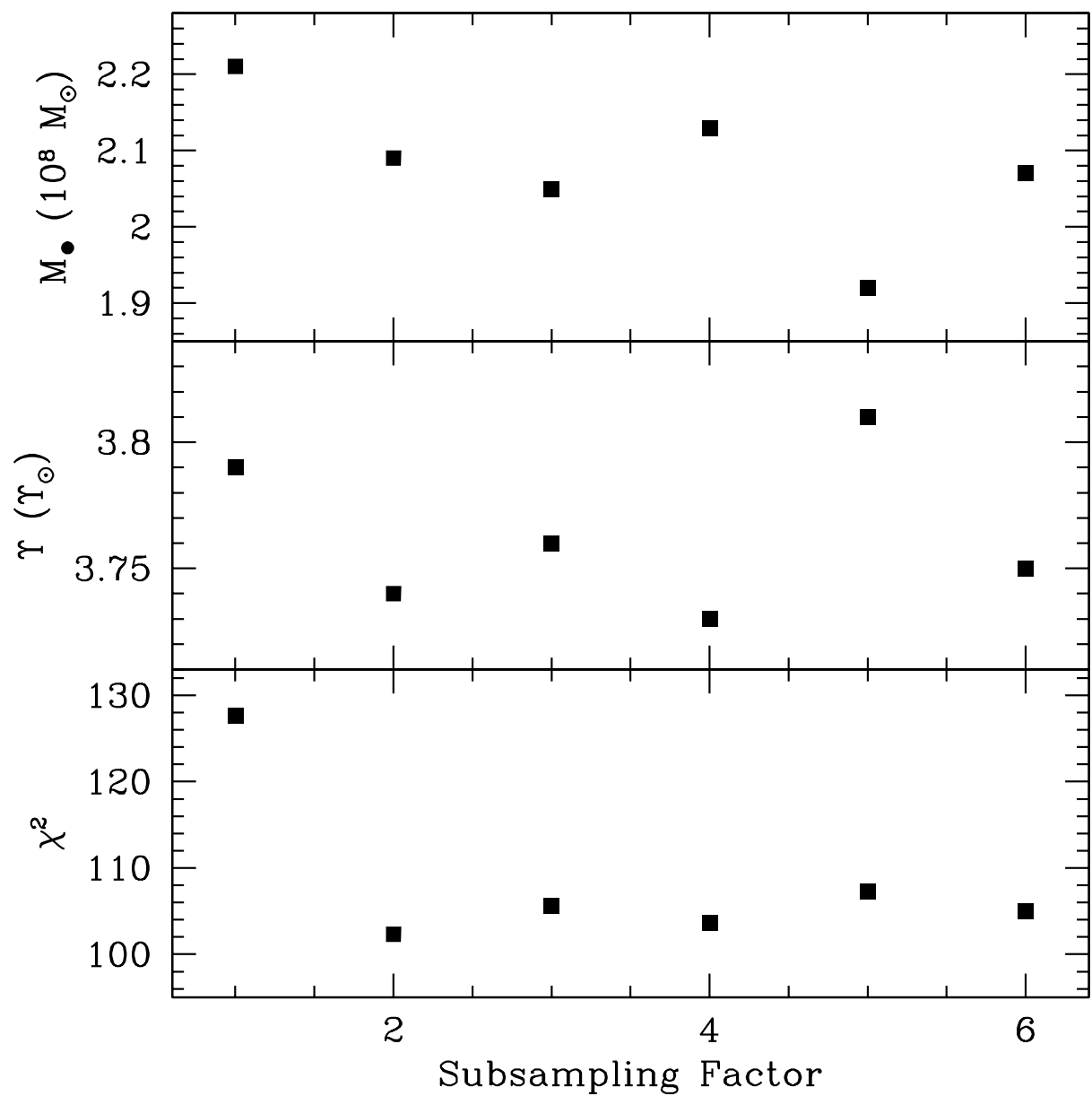}
\caption{Results of model fits for different values of the subsampling
factor $s$.  The quantities \mbh, \ml, \inc, $\theta$, and \vsys\ were
allowed to float as free parameters.  These models were calculated
with no asymmetric drift correction, and with \rfit\ = 0\farcs5.
\label{figsubsamp}}
\end{center}
\end{figure}

To optimize the model fits, we performed \chisq\ minimizations using
the downhill simplex algorithm described by \citet{nr}.  Initially,
the five parameters \mbh, \ml, \inc, $\theta$, and \vsys\ were allowed
to vary independently in the minimization.  We first discuss models
without a correction for asymmetric drift.  The preliminary models
were calculated out to \rfit\ = 0\farcs5, giving a
$1\arcsec\times1\farcs2$ rectangular region.  To test the effects of
the grid sampling size, we ran models for values of $s$ from 1 to 6.
Different choices for $s$ lead to very similar results (Figure
\ref{figsubsamp}), with a root-mean-square (RMS) scatter in \mbh\ of
only $4\%$ for $s$ in the range 1--6.  The models with $s=2$ yield the
best fits, and we performed our final calculations using $s=2$.  For
$\rfit = 0\farcs5$ and $s=2$, there are 105 points in the velocity
curves and the total \chisq\ of the best-fitting model is 102.3.

\begin{figure}
\begin{center}
\includegraphics{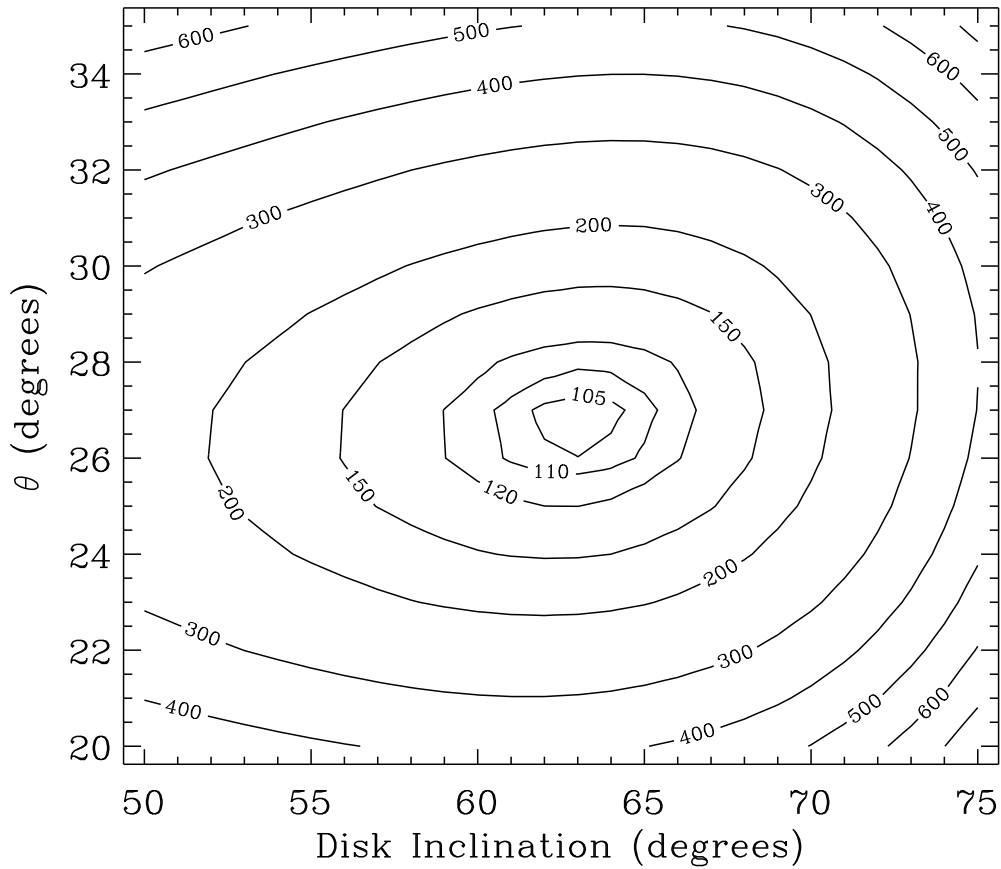}
\caption{Contours of constant \chisq\ for models with varying values
of disk inclination $i$ and offset angle $\theta$ between the slit
P.A. and the disk major axis.  Models were calculated with no
asymmetric drift correction, $s=2$, and \rfit\ = 0\farcs5.  At each
grid point, \mbh, \ml, and \vsys\ were allowed to float freely.
\label{contours}}
\end{center}
\end{figure}

After many trial runs with different initial parameter values, we
found that the model fits consistently converged on $\inc = 63\arcdeg$
and $\theta = 27\arcdeg$.  These values are consistent with
expectations from the WFPC2 imaging data.  To explore the range of
uncertainty on the disk orientation parameters, we calculated a grid
of models over a range of values of \inc\ and $\theta$, leaving \mbh,
\ml, and \vsys\ as free parameters in the fits.  Figure \ref{contours}
displays contours of constant \chisq\ for this model grid,
demonstrating that with five parallel slit positions the gas
kinematics can tightly constrain the disk orientation parameters (to
better than $\pm1\arcdeg$ at 68.3\% confidence).  

At radii beyond the edge of the disk, the ionized gas velocities begin
to deviate systematically from circular motion.  Beyond $\sim150$ pc
from the nucleus, the WFPC2 F547M image shows a filamentary structure
in dust absorption which does not have obvious circular symmetry, and
we do not attempt to fit kinematic models to this region.

The galaxy's systemic velocity \vsys\ was left as a free parameter and
determined separately for each model calculation.  From the
best-fitting model, we find \vsys\ = 1388 \kms.  Different model runs
with reasonable parameters yielded results within a few \kms\ of this
value.  \citet{huc90} list a value of $1358\pm15$ \kms\ for the
recession velocity of NGC 3245.  The discrepancy is most likely caused
by the asymmetric distribution of emission-line surface brightness
across the disk, which would bias a ground-based measurement toward
the lower velocities of the northern side of the disk.  For \vsys\ =
1388 \kms, our model predicts a surface-brightness-weighted mean
velocity of 1360 \kms, in good agreement with the ground-based
redshift.

In all of the calculations described above, the center of rotation was
assumed to be at the same location as the peak of the \hn\ surface
brightness at the nucleus.  As a test, we computed models in which the
projected velocity field map was shifted in $x$ and $y$ relative to
the emission-line surface brightness map.  We found that shifting the
velocity field by even one STIS pixel in any direction (i.e., $\sim5$
pc) led to significant increases in \chisq, and we conclude that the
dynamical center is effectively coincident with the optical peak in
the WFPC2 \hn\ image.

\subsection{Fitting Region and Best-Fitting Models}

\begin{figure}
\begin{center}
\plotone{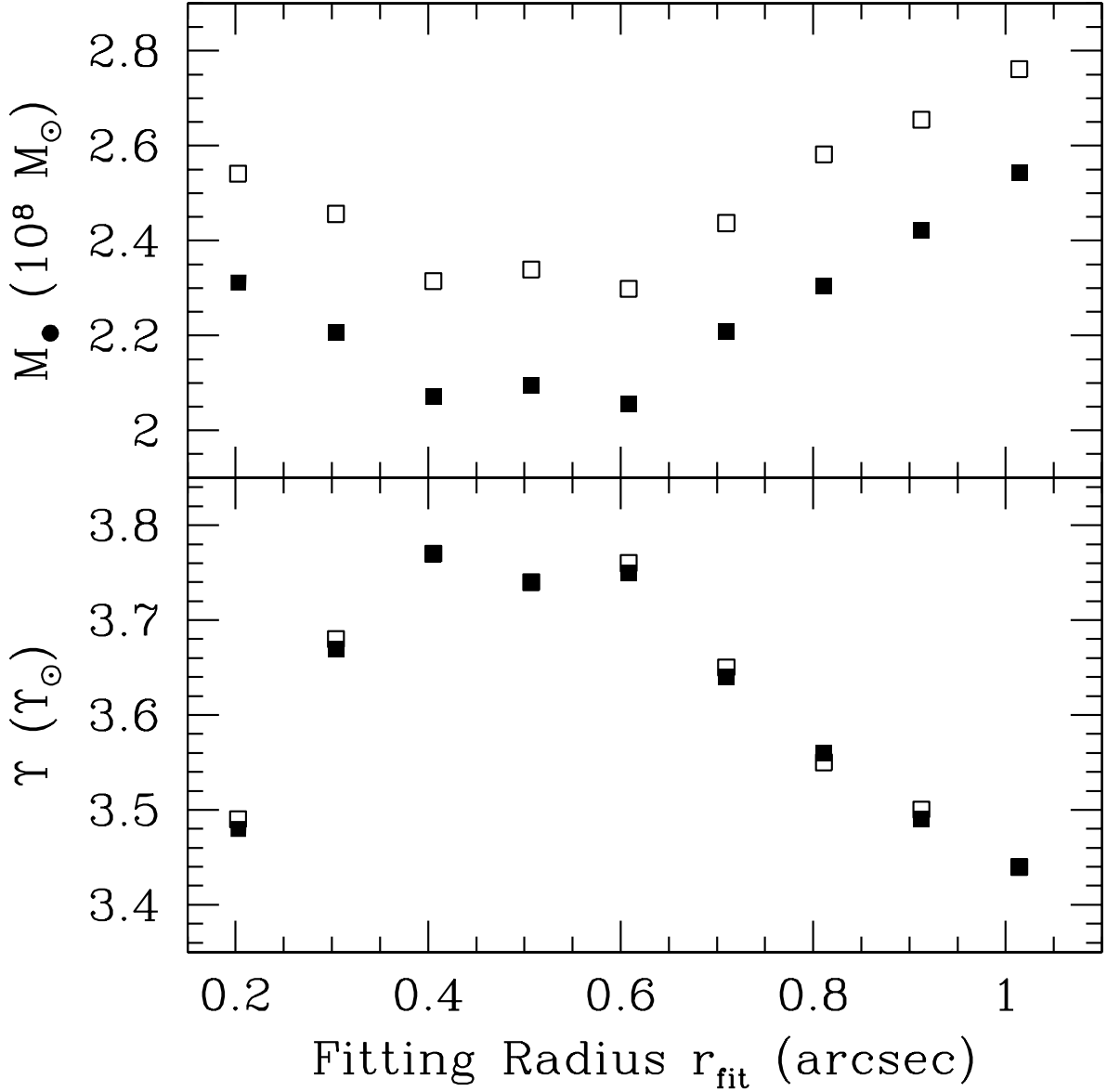}
\caption{Results of model fits for different values of \rfit.  In each
calculation, \mbh\ and \ml\ are allowed to float while \inc\ and
$\theta$ are fixed at 63\arcdeg\ and 27\arcdeg, respectively.  Models
with and without the asymmetric drift correction are indicated by open
and filled squares, respectively.
\label{figradius}}
\end{center}
\end{figure}

\begin{figure}
\scalebox{0.7}{\includegraphics{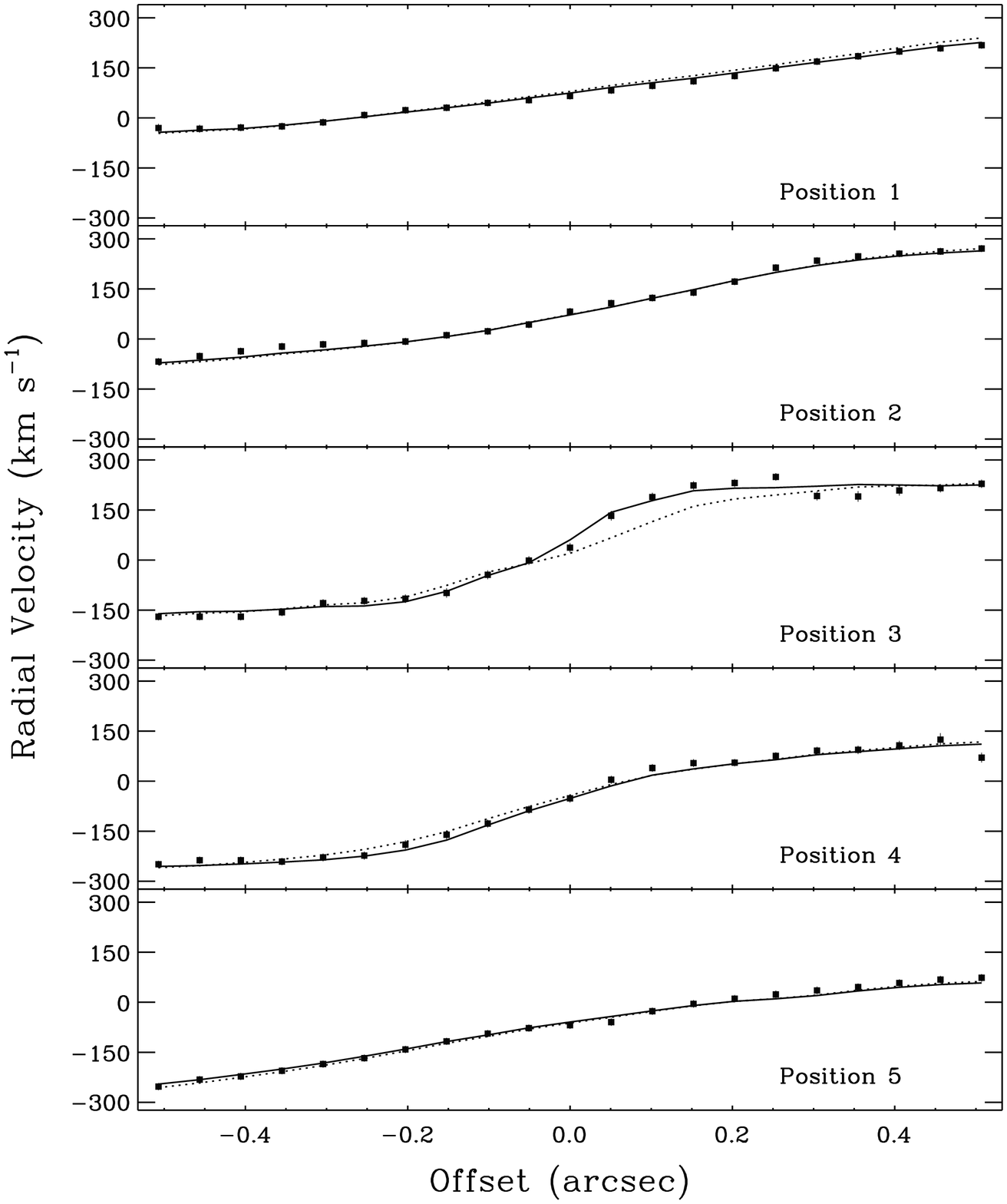}}
\caption{Comparison of the \hn\ radial velocities measured from
the STIS data with model predictions.  \emph{Solid curve:}
Best-fitting model with \rfit\ = 0\farcs5.  This model has
$\mbh=2.09\times10^8 \msun$ and $\ml = 3.74 \mlsun$.  \emph{Dashed
curve:} The best-fitting model for the case \mbh\ = 0 and \rfit\ =
0\farcs5.  For this model, \ml\ = 4.61 \mlsun.
\label{innerrotcurves}}
\end{figure}

\begin{figure}
\scalebox{0.7}{\includegraphics{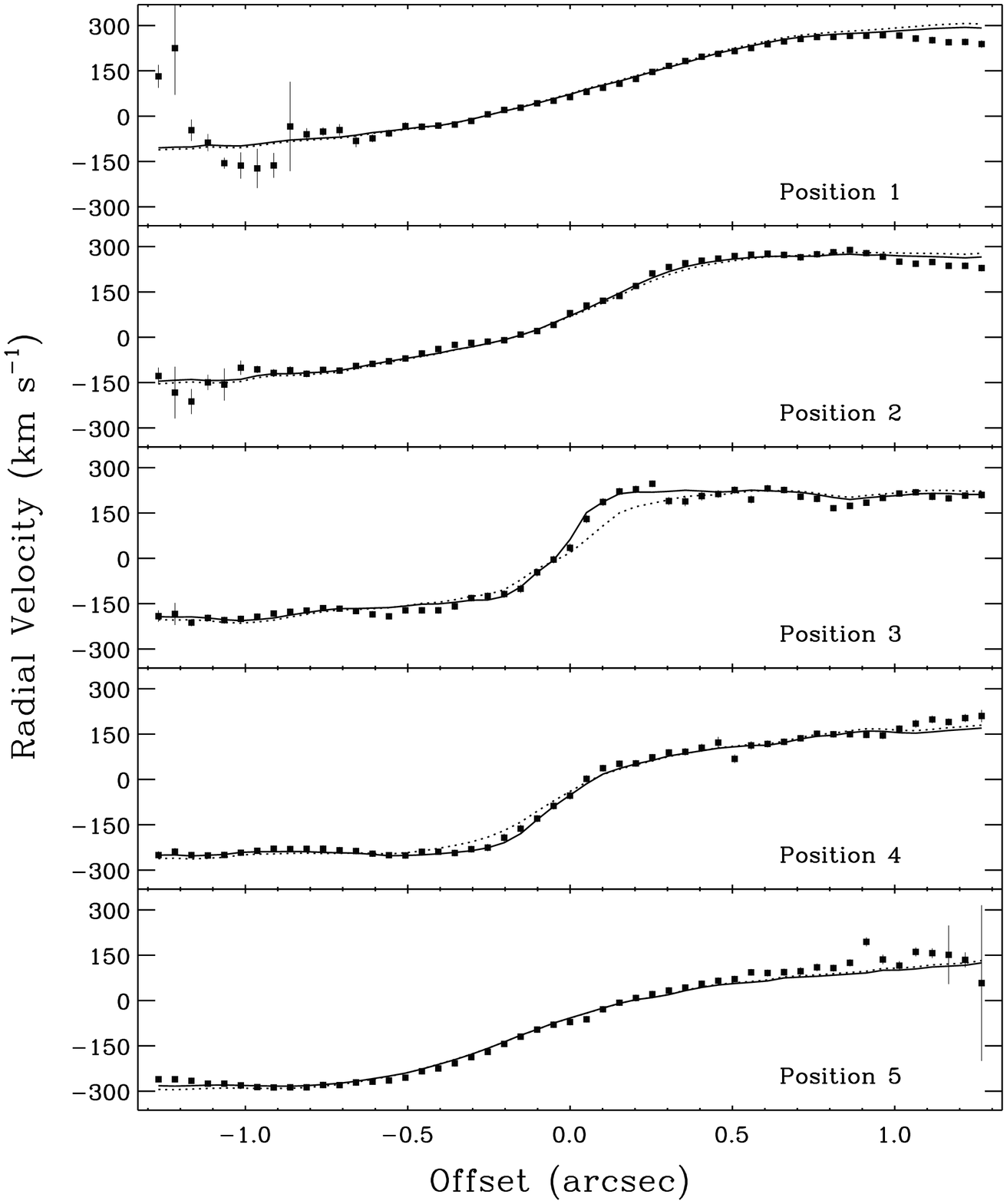}}
\caption{Comparison of the \hn\ radial velocities measured from the STIS
data with model predictions.  \emph{Solid curve:} Best-fitting model
for the case \rfit\ = 1\farcs0.  This model has $\mbh=2.54\times10^8
\msun$ and $\ml = 3.44 \mlsun$.  \emph{Dotted curve:} Best-fitting
model for the case \mbh\ = 0 and \rfit\ = 1\farcs0.  This model has
\ml\ = 4.04 \mlsun.
\label{outerrotcurves}}
\end{figure}

The model fit with $s=2$ and $\rfit=0\farcs5$ gives a best-fit BH mass
of $2.09\times10^8$ \msun\ and $\ml = 3.74$ in $R$-band solar units.
One way to gauge the uncertainty in the results is to run models with
a range of values of \rfit.  Fixing \inc\ = 63\arcdeg\ and $\theta =
27\arcdeg$, we ran model fits with \rfit\ ranging from 0\farcs2 to
1\farcs0.  The results, shown in Figure \ref{figradius}, demonstrate
that different choices of \rfit\ can result in values of \mbh\ in the
range $(2.0-2.6)\times10^8$ \msun, for models with no asymmetric
drift.  Figures \ref{innerrotcurves} and \ref{outerrotcurves} show the
best-fitting models for \rfit\ = 0\farcs5 and \rfit\ = 1\farcs0, and
also the best-fitting models for the case \mbh\ = 0.  

For large values of \rfit\ (greater than about 0\farcs5), \chisq\ is
dominated by points far from the BH, and the models have relatively
little sensitivity to \mbh.  Also, with $\rfit > 0\farcs5$ there are
data points outside the optical edge of the disk, and these points
appear to deviate systematically from any circular rotation model,
particularly in slit position 5.  The deviations from model
predictions outside the main body of the disk are a possible sign of a
warp, or the kinematic signature of gas which has not yet settled into
the disk.  Thus, the derived BH mass probably becomes unreliable as
\rfit\ increases beyond 0\farcs5.

Even in the inner portions of the disk, some points in the velocity
curves deviate from the circular rotation models.  For example, there
is a ``step'' in the velocity curve for slit position 3 at 0\farcs3
south of the nucleus, and a region 0\farcs3--0\farcs6 north of the
nucleus has a rotation speed too large to be fit by the models.  The
influence of these regions on the model fits depends on the size of
the fitting region, and partly explains the variation in the derived
value of \mbh\ for varying values of \rfit.

The dependence of \mbh\ on \rfit\ may also result at least in part
from a real gradient in the $R$-band mass-to-light ratio across the
inner arcsecond of the galaxy.  As shown by \citet{car97}, early-type
galaxies often have radial variations in $V-I$ color of 0.1--0.2 mag
over the inner arcsecond.  An intrinsic color gradient in NGC 3245
(presumably due to an underlying metallicity gradient) could
conceivably lead to variations in \ml\ at the level observed in NGC
3245.  Another possibility is that the inner disk may be slightly
warped, so the mass derived from fitting planar disk models would
depend on the fitting region.  

We choose $\rfit = 0\farcs5$ for defining our ``best-fit'' value of
\mbh, as this region is small enough to be entirely contained within
the disk, but large enough to contain a sufficient number of data
points that the model fits can be constrained tightly.  In principle,
one could attempt to select an ``optimal'' fitting region by finding
the value of \rfit\ that would lead to the minimum value of \chisqdof.
However, for this dataset we find that the minimum value of \chisqdof\
occurs at the smallest possible values of \rfit.  As \rfit\ increases,
\chisqdof\ increases due to the influence of discrepant regions which
depart slightly from circular rotation.  By choosing \rfit\ large
enough to cover much of the disk, we ensure that any individual
discrepant region will not have a substantial effect on the results.
We note also that models fitted to only the three central slits result
in values of \mbh\ which differ by $\lesssim5\%$ from the values
derived using all five slits.

\begin{figure}
\plotone{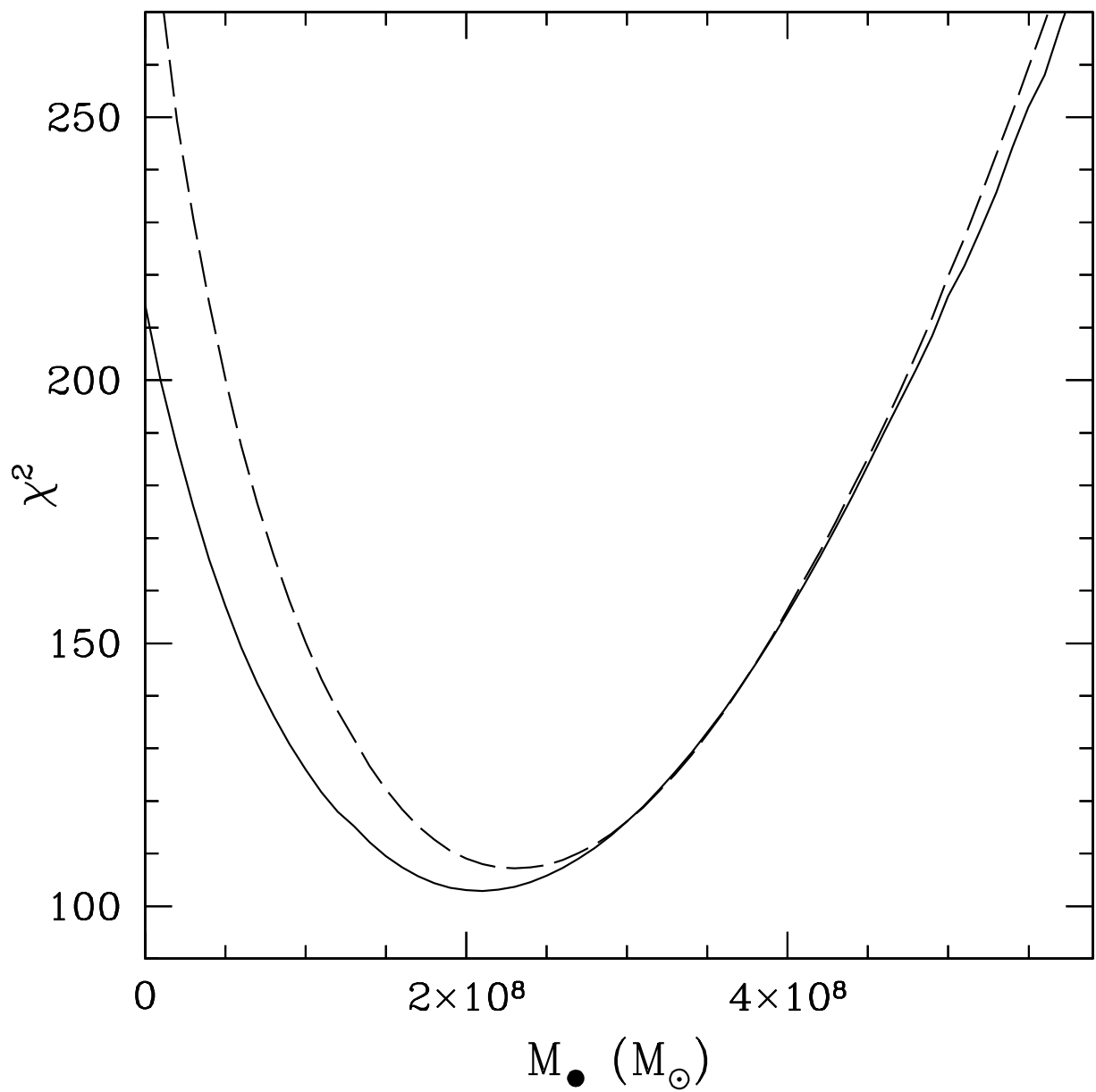}
\caption{Results of disk-model fits for \rfit\ = 0\farcs5 and $s=2$.
Each model was calculated by fixing \mbh\ and then optimizing \ml,
\inc, $\theta$, and \vsys\ to minimize \chisq\ for the given value of
\mbh.  \emph{Solid curve:} Models without asymmetric drift correction.
\emph{Dashed curve:} Models including the asymmetric drift correction.
\label{figchisq}}
\end{figure}

Figure \ref{figchisq} shows the results of models calculated for
\rfit\ = 0\farcs5, for \mbh\ ranging from 0 to $6\times10^8$ at
intervals of $0.1\times10^8$ \msun.  At each fixed value of \mbh, the
parameters \ml, \inc, $\theta$, and \vsys\ were left as free
parameters.  For models without asymmetric drift, \mbh\ is constrained
to lie within $(1.8-2.3)\times10^8$ \msun\ at 68.3\% confidence, and
within $(1.5-2.7)\times10^8$ \msun\ at 99\% confidence, with a
best-fitting value of $2.1\times10^8$ \msun.  Models without a central
dark mass are ruled out at better than 99.99\% confidence.  With the
asymmetric drift correction included, the best-fitting BH mass
increases to $2.3\times10^8$ \msun, with 68.3\% and 99\% confidence
ranges of $(2.1-2.5)\times10^8$ and $(1.8-2.9)\times10^8 \msun$,
respectively.  Including the asymmetric drift correction increases the
minimum value of \chisq\ from 102.3 to 106.9.  Since the models
without asymmetric drift provide better fits, we use these models to
derive the BH mass, but we allow for the possibility of asymmetric
drift by including it in the estimated uncertainty range for \mbh.

The \chisq\ curve appears smooth despite the fact that the models are
calcualted on a discrete spatial grid, because the values of \mbh\ and
\ml\ are continuous variables and \chisq\ can vary smoothly as these
parameters change.  We also note that the orientation parameters \inc\
and $\theta$ are so tightly constrained that the curves of minimum
\chisq\ in Figure \ref{figchisq} would appear nearly unchanged if
these parameters were held fixed at their best-fitting values rather
than being allowed to vary in the fits.

Over the range $\rfit=0\farcs2-0\farcs5$, the derived value of \ml\ is
3.5--3.8 in $R$-band solar units.  This is comparable to $R$-band
mass-to-light ratios observed in other early-type galaxies.  For a
sample of ellipticals, \citet{vdm91} found a mean value of $\Upsilon_R
= (3.32 \pm 0.14) h_{50}$, where $h_{50}$ is the Hubble constant in
units of 50 km s\per\ Mpc\per.

\subsection{Error Budget}
\label{errorbudget}

We summarize the sources of uncertainty as follows.

\emph{Stellar profile fitting:} We investigated the uncertainty in
\mbh\ due to the uncertainty in deprojecting the galaxy's surface
brightness profile by running models in which the stellar circular
velocity curve \vstar\ was displaced by $\pm1\sigma$ from its default
value.  The confidence limits on \vstar\ were determined by
deprojecting different Monte Carlo realizations of the surface
brightness profile, on the basis of the $1\sigma$ uncertainty of each
$R$-band isophote from the surface brightness profile fit.  A total of
100 realizations of the surface brightness profile were used to derive
the uncertainty on \vstar.  For models with increased or decreased
stellar surface brightness, the resulting change in \mbh\ is
(respectively) $-5.7\%$ or $+1.9\%$.

\emph{Model fitting uncertainty:} The 68.3\% confidence interval on
\mbh\ (corresponding to a $1\sigma$ uncertainty) from the model
fitting is $(1.8-2.3)\times10^8$ \msun, for models without the
asymmetric drift correction.  This includes the (negligible)
uncertainty due to the disk-orientation parameters \inc\ and $\theta$,
which were allowed to vary in the model fits.

\emph{Asymmetric Drift:} The asymmetric drift correction raises the
best-fitting BH mass by $0.25\times10^8$ \msun, or 12\%.  However, it
is not clear whether there actually is asymmetric drift in the NGC
3245 disk.  To account for the possibility that there may be
asymmetric drift, we propagate a 12\% uncertainty into the upper error
bar on \mbh.

\emph{Distance:} The mass derived from modeling the velocity field is
proportional to $D$.  \citet{ton01} give an uncertainty of 0.20 mag in
the distance modulus to NGC 3245, corresponding to a 10\% uncertainty
in distance.  In addition, there is a systematic uncertainty of 9\% in
the Cepheid-based zeropoint of the distance scale \citep{mou00}.  We
propagate only the random error in the distance into our result, since
changes in the zeropoint of the distance scale will affect all BH mass
measurements equally.

\emph{Fitting radius:} A major source of uncertainty in \mbh\ is the
choice of fitting radius.  For \rfit\ in the range
$0\farcs2-0\farcs5$, \mbh\ varies between $2.0\times10^8$ and
$2.3\times10^8$ \msun, and we propagate this range into the $1\sigma$
uncertainty on \mbh.  As \rfit\ increases beyond 0\farcs5, \mbh\ rises
to $\geq2.4\times10^8$ \msun, but for such large fitting regions the
results are unreliable because of noncircular motions outside the
disk, and because the fit is dominated by points far from the BH.

\emph{Subsampling factor:} For \rfit\ = 0\farcs5, models computed with
different values of $s$ have an RMS scatter of $9\times10^6$ \msun, or
4\% of the best-fitting BH mass. This variation results from the
graininess of the calculation and the differences in the \hn\
surface-brightness model for different subsampling factors.  We
propagate this RMS scatter into our final uncertainty range.

\emph{Differences between \hal\ and [\ion{N}{2}] kinematics:} Our
measurements of the STIS data give a mean velocity from the \hal\ and
[\ion{N}{2}] emission lines together, and we have not attempted to
measure or fit models to the individual velocity curves of \hal\ and
[\ion{N}{2}].  As discussed in \S\ref{sectionlinewidth}, these two
lines have different widths, and the models described above were
calculated using the intrinsic velocity dispersion that provided the
best fit to the [\ion{N}{2}] data.  We also ran a model in which the
intrinsic velocity dispersion of the gas was set to match \hal\ rather
than [\ion{N}{2}], to test whether this would change the result
substantially.  The model fit converged on $\mbh = 1.91\times10^8$
\msun, or 9\% lower than the best-fitting mass for our standard
models.  To account for the possibility of differences in the
kinematics of different ionic species, we propagate an uncertainty of
9\% into our final result.  

Combining all of these sources of uncertainty, we arrive at a final
range of $\mbh = (1.6-2.6) \times 10^8$ \msun, with a best-fit value
of $2.1\times10^8$ \msun.  This mass range is meant to correspond to a
$1\sigma$ confidence limit, although the uncertainty range due to the
choice of fitting radius is somewhat subjective and cannot be
expressed in terms of formal confidence intervals.

One additional potential source of uncertainty is the correction for
dust extinction in the $R$-band image.  Had we measured the stellar
profile without correcting for extinction at all, the derived BH mass
would have been $2.6\times10^8$ \msun.  While we do not attempt to
quantify the uncertainty in the extinction correction, the fact that
the entire correction only changes the BH mass by less than 25\%
suggests that the uncertainty in the correction will not be a
significant contributor to the overall error budget.

\subsection{How Compact is the Central Dark Mass?}

\begin{figure}
\plotone{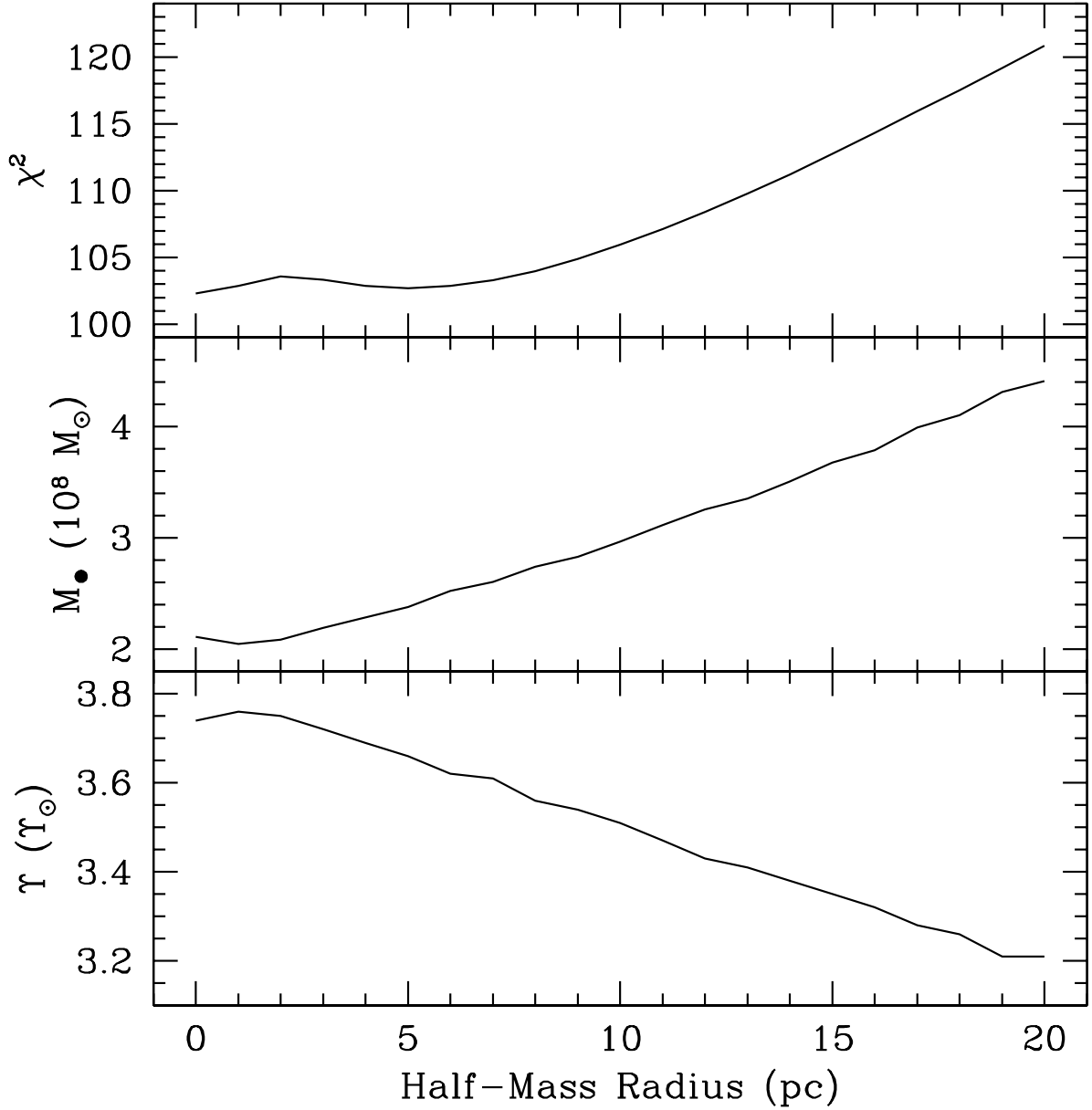}
\caption{Results of model fits for models with a spatially extended
central dark mass, modeled as a Plummer sphere.  For each fixed value
of \rhalf, models were calculated with the quantities \mbh, \ml, and
\vsys\ as free parameters.  
\label{figplummer}}
\end{figure}

Even if a central dark mass is detected, our observations cannot
determine whether it is truly a single compact object.  To test the
possible spatial extent of the dark mass, we computed models using a
Plummer model to represent the density distribution of the dark mass
\citep[e.g.,][]{mac97, mao98}.  This model is an extreme case in the
sense of having a very low central density for a given half-mass
radius.  For a given central mass, the rotation curves for these
extended-mass models are identical at large radii to the models with a
point-like central mass, but at small radii the central slopes are
shallower.  We ran model fits for values of the half-mass radius
(\rhalf) of the Plummer model ranging from 0 to 20 pc at intervals of
1 pc (Figure \ref{figplummer}).  The models were run with $s=4$ to
improve the spatial sampling at the nucleus.

The best-fitting models are obtained for a pointlike central mass.  As
\rhalf\ increases beyond 1 pc, the total dark mass required to fit the
central velocity slope rises correspondingly.  The increase in
\chisq\ indicates that the central mass must have $\rhalf < 7$ pc at
68.3\% confidence, or $\rhalf < 12$ pc at 99\% confidence.  The
corresponding density of the dark mass must be $>1.0\times10^5$ \msun\
pc\percu\ at 68.3\% confidence, or $>2.3\times10^4$ \msun\ pc\percu\
at 99\% confidence, within the half-mass radius.  This is comparable
to the central densities found in most other BH hosts, but
significantly lower than the inferred central densities for galaxies
such as M87 in which the BH sphere of influence is better resolved
(see Maoz 1998).

\subsection{Comparison of Models with Different Features}

To determine whether the level of detail in our modeling code actually
provides a noticeable improvement in the final results, we also ran a
series of models where various features of the code were ``turned
off.''  We calculated models in which the \hn\ surface brightness was
set to a constant, or set to the exponential surface-brightness model;
models in which the PSF was set to a delta function at the limit of
the model resolution; models in which the instrumental wavelength
shifts of light entering the slit off-center were omitted; and models
in which the STIS pixels were not subsampled.  In all of these
calculations, we used \rfit\ = 0\farcs5.

As shown in Table \ref{table2}, the instrumental wavelength shifts
alone do not result in a substantial change in the derived BH mass,
but including them does improve the fits to the observed rotation
curves.  Neglecting the PSF structure can lead to dramatically
different model results, however.  In general, models which do not
include blurring effects (the PSF and nonzero aperture size) will tend
to underestimate the BH mass because a properly blurred model will
require a greater central mass to match a given rotation velocity.

\begin{figure}
\scalebox{0.7}{\includegraphics{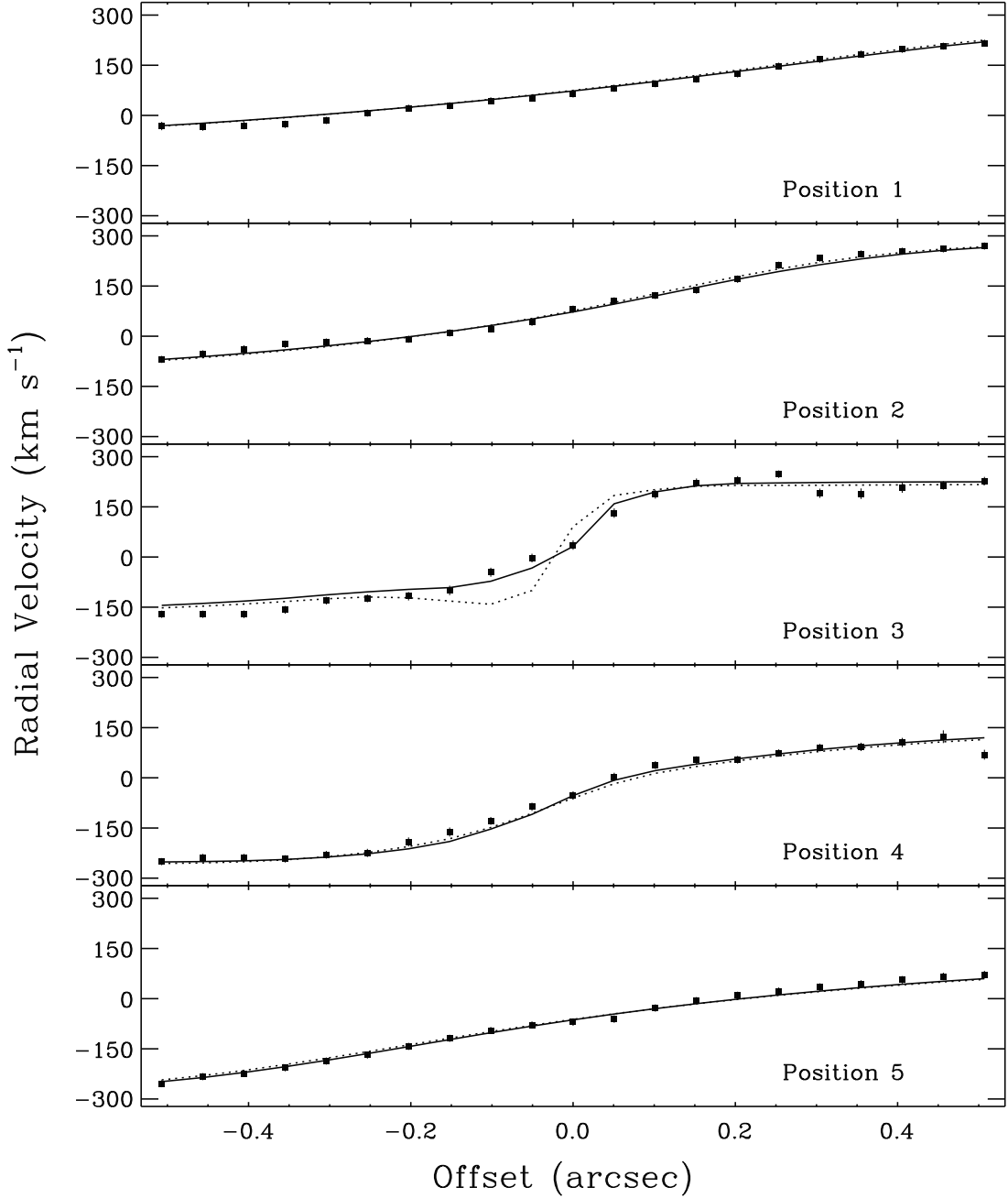}}
\caption{Result of model calculations using the exponential
surface-brightness model.  \emph{Solid curve:} Results for $s=1$.
\emph{Dotted curve:} Results for $s=4$.  Both models are computed with
Tiny Tim PSFs and with instrumental velocity shifts; parameters are
given in Table \ref{table2}.
\label{crappymodel}}
\end{figure}

The model results are also sensitive to the treatment of the
emission-line surface brightness.  In particular, use of the
exponential surface-brightness model led to difficulties.  For $s=1$,
the model velocity curves appear to fit the data adequately (Figure
\ref{crappymodel}), but at higher values of $s$, the increased
peakiness of the central surface brightness led to large velocity
deviations in the innermost regions of slit position 3.  For high
values of the subsampling factor, reasonable results could only be
obtained if the exponential surface-brightness profile was assumed to
flatten out in the inner core, or if there was an inner ``hole'' in
the \hn\ distribution. This problem is not unique to the exponential
model; it would occur for any other model with a comparably steep
profile at the nucleus.  The \hn\ surface brightness model derived
from the WFPC2 image does not suffer from this problem because its
nuclear profile is not nearly as sharply peaked.  With the exponential
surface-brightness model and $s=1$, the best-fitting BH mass is
$1.5\times10^8$ \msun, significantly below the best-fitting result for
models incorporating the actual \hn\ surface-brightness distribution.
The radial velocity curves in Figure \ref{figmiddle} also illustrate
the importance of including an accurate model for the
surface-brightness distribution.

Notably, the lowest value of \chisq\ was found for the ``full'' models
which included instrumental wavelength shifts, the Tiny Tim PSF, and
the actual \hn\ surface brightness derived from the WFPC2 F658N image.
None of these features adds any free parameters to the fit, so the
decrease in \chisq\ represents a real and significant improvement in
model results when all available information is taken into account in
the modeling code.  In comparison with our best-fitting model, all
models using the exponential surface-brightness profile are excluded
at $>99.99\%$ confidence.

These results underscore the need for accurate modeling in future
gas-dynamical BH measurements.  The narrow-band image can be a
particularly important input to the models, especially for galaxies
such as NGC 3245 in which the emission-line surface brightness is
patchy or irregular.  Using the 0\farcs1-wide slit would alleviate
these problems to some extent, but at the expense of reduced S/N.
Also, we note that many of the features in our modeling procedure
should be equally applicable to stellar-dynamical BH measurements.
Instrumental wavelength shifts and separation of the PSF into optical
and charge-diffusion components will affect the line-of-sight velocity
profiles and velocity dispersions derived from stellar absorption
lines, but stellar-dynamical BH measurements done with \hst\ data have
generally not treated instrumental effects at this level of detail.

\section{Discussion}
\label{discussion}

\subsection{Is the Central Dark Mass a BH?}

With numerous recent claims of detections of BHs in galactic nuclei
based on \hst\ data, it is important to remember that these
observations cannot prove that a BH has actually been detected.
Strictly speaking, we have detected a massive dark object and
determined a lower limit to its density.  At best, our data probe
scales of $\sim10^5$ Schwarzschild radii, and at this resolution we
cannot definitively rule out alternatives to the BH interpretation.
\citet{mao98} has shown that a hypothetical cluster of dark objects
such as stellar-mass BHs, neutron stars, brown dwarfs, or
planetary-mass objects could in principle have an extremely long
lifetime against evaporation or collisions.  The nuclei of our Galaxy
and NGC 4258 are the only two cases where the inferred central
densities are sufficiently great that any hypothetical dark cluster
would have a lifetime shorter than the Hubble time.  While such ``dark
cluster'' models may not seem plausible or well motivated, they are
also not ruled out by the available data.  For most claimed BH
detections including NGC 3245, the conclusion that the massive dark
object is a BH still rests on circumstantial evidence and on the
expectation that a supermassive BH ought to be present in the nuclei
of most massive galaxies \citep[e.g.,][]{sol82}.

\subsection{Asymmetric Drift and the Intrinsic Velocity 
Dispersion}

The origin of the intrinsic velocity dispersion in nuclear gas disks
remains a puzzling issue, particularly in light of the differing
velocity dispersions of \hal\ and [\ion{N}{2}] in NGC 3245.  The fact
that the [\ion{N}{2}]/\hal\ intensity ratio is highest in regions
having the highest velocity dispersion suggests that shock excitation
may be responsible for the LINER emission near the center of the disk.
This possibility could be tested with observations of a wider range of
emission lines.  It is reassuring that the asymmetric drift correction
does not lead to a very substantial change in \mbh\ for NGC 3245 when
calculated for [\ion{N}{2}]; for \hal\ the inferred asymmetric drift
would be an even smaller effect.  However, the effect may be more
important in other galaxies.  In IC 1459 the gas velocity dispersion
rises to $\sim500$ \kms\ at the nucleus, and if there is asymmetric
drift then the BH mass could be a factor of $\sim3-4$ greater than the
mass derived from cold disk models \citep{vk00}.  This does make the
BH detection more secure, but raises the uncertainty on \mbh\ far
beyond the level of the uncertainty in the model fitting.  The nuclear
disk in M87, on the other hand, does not appear to have a substantial
intrinsic velocity dispersion relative to the circular velocity
\citep{mac97}.

It appears that asymmetric drift may play an important role in the
dynamical structure of some, but perhaps not all, nuclear gas disks.
With new results from the various STIS programs currently in progress,
it should be possible to determine whether there is any relation
between host galaxy properties and the intrinsic velocity dispersion
in nuclear disks.  For example, is the ratio $\sigma / v_c$ in nuclear
disks correlated with \mbh\ or with bulge properties?  Also, can any
galaxies be found in which an asymmetric drift correction improves the
model fits?  If asymmetric drift does turn out to be an important
effect in many nuclear disks, then some previous BH mass measurements
might have underestimated the true BH masses by a substantial factor.

\subsection{Host Galaxy Properties} 

\citet{geb00} and \citet{fm00} \citep[see also][]{mf00} have shown
that there is a very small scatter in the relation between \mbh\ and
bulge velocity dispersion, at least for the galaxy samples they
studied.  Having additional BH mass measurements makes it possible to
refine and test these correlations.  For NGC 3245, the central
velocity dispersion ($\sigma_c$) is 230 \kms\ \citep{td81}.  The
relation derived by \citet{mf00} predicts $\mbh = 2.5\times10^8$
\msun\ for this velocity dispersion.  The $\mbh - \sigma$ relation
derived by \citet{geb00} uses the surface-brightness weighted velocity
dispersion within the effective radius ($\sigma_e$), and we do not
have a measurement of this quantity.  However, as \citet{geb00} show,
the difference between $\sigma_e$ and $\sigma_c$ is typically smaller
than 10\%.  Assuming that $\sigma_e$ is close to 230 \kms, the
\citet{geb00} relation predicts $\mbh = 2.0 \times 10^8 \msun$, with
an uncertainty of $\sim30\%$ due to the possible difference between
$\sigma_e$ and $\sigma_c$.  Our results further support the predictive
power of the $\mbh-\sigma$ relation, as both the Gebhardt \etal\ and
the Merritt \& Ferrarese relations predict values of \mbh\ which are
within the $1\sigma$ uncertainty range for NGC 3245.

NGC 3245 has an extinction-corrected absolute magnitude of $M_B =
-20.08$ \citep{hfs97a}, which corresponds to a $B$-band luminosity of
$1.5\times10^{10}$ \lsun.  The bulge accounts for 76\% of this total
$B$-band luminosity, as determined by using the bulge-disk
decomposition parameters from \citet{bba98} to subtract off the disk
contribution.  Assuming a typical $B$-band mass-to-light ratio of 6
\citep{vdm91}, the bulge mass is $\sim 6.8 \times 10^{10}$ \msun.
Then, $\mbh / \mbul \approx 0.003$.  The median $\mbh/\mbul$ seen in
other galaxies is $\sim0.002$, but there is nearly an order of
magnitude of scatter about this best-fit ratio \citep{ho99}.

\section{Conclusions}
\label{sectionconclusions}
 
We have detected a massive dark object, presumably a BH, in the
nucleus of NGC 3245.  Its mass is constrained to be in the range
$(1.6-2.6)\times10^8$ \msun, and the $R$-band mass-to-light ratio is
in the range 3.5--3.8 in solar units.  Models without a central dark
mass predict radial velocity curves which are too shallow to match the
data in the innermost regions, and are excluded at a formal confidence
level better than 99.99\%.  The STIS data provide excellent spatial
coverage of the nuclear disk and high S/N, so we are able to verify
that the disk kinematics are well matched by circular rotation models.

Our results demonstrate the importance of modeling the effects of the
telescope and spectrograph optics, so that dynamical models for the
rotating gas disk can be adequately compared with the data.  Similar
issues have also been discussed by \citet{vv98} and \citet{mb01}.
Models which do not take into account the nonzero aperture size or the
PSF may be prone to unpredictable errors.  Modeling the apparent
wavelength shifts for light entering the slit off-center does not lead
to significant changes in the derived value of \mbh, but it does
improve the quality of the model fits, and allows the models to make
detailed predictions for the line widths.  A new aspect of our
modeling technique is the use of the measured surface brightness
distribution of emission-line light in the calculations, rather than a
smooth approximation.  Folding the WFPC2 narrow-band image into the
models improves the quality of the fits significantly, and
demonstrates that some of the apparent irregularities in the velocity
field are actually due to local variations in surface brightness
rather than local departures from circular rotation.  We have also
included other minor effects such as anamorphic magnification in the
spectrograph optics and the separation of the PSF into optical and
charge-bleeding components.  While these do not alter the results
dramatically, they are real effects which do occur in the spectrograph
and they can easily be modeled without increasing the computation
time.

For future progress in gas-dynamical BH mass measurements, careful
choice of targets is crucial, because fewer than 20\% of disk galaxies
have emission-line velocity fields which are amenable to this type of
analysis \citep{sar01}.  On the theoretical front, the question of the
dynamical state of these nuclear disks and the possible role of
asymmetric drift remain important unresolved issues.  Also, variations
in the kinematics of different emission lines should be investigated
further.  Despite these difficulties, our results demonstrate that the
gas-dynamical method can be used to measure BH masses to better than
25\% accuracy with \hst\ STIS data.  We plan to apply these techniques
to a larger sample of galaxies in the future, as a step toward filling
in the $\mbh-\sigma$ correlation over the entire mass range of BHs in
galactic nuclei.

\acknowledgments

Support for this work was provided by NASA through grant number
GO-07403 from STScI, which is operated by AURA, Inc., under NASA
contract NAS 5-26555.  NASA grant NAG5-3556 is also acknowledged.
Research by A.J.B. is supported by a postdoctoral fellowship from the
Harvard-Smithsonian Center for Astrophysics.  We thank Walter Dehnen,
Krzysztof Stanek, and Roeland van der Marel for illuminating
discussions, and John Tonry for providing the SBF distance to NGC 3245
prior to publication.  A.J.B. is grateful to the Max-Planck-Institut
f\"ur Astronomie for its hospitality during a visit in September 2000.
We thank the referee, Gary Bower, for several helpful suggestions
which improved the presentation of this paper.


\begin{center}
\begin{deluxetable}{lcrc}
\tablewidth{3.7in}
\tablecaption{Summary of STIS Observations\label{table1}}
\tablehead{\colhead{Slit} & \colhead{\hst\ Image} & \colhead{Offset} &
\colhead{Exposure Time} \\
\colhead{Position} & \colhead{Rootname} & \colhead{(arcsec)} &
\colhead{(s)}}
\startdata
1 & o57205010 & 0.50 & 2961 \\
2 & o57205020 & 0.25 & 2715 \\
3 & o57205030 & 0.00  & 2715 \\
4 & o57205040 & $-0.25$  & 2715 \\
5 & o57205050 & $-0.50$ & 4651 \\
\enddata

\tablecomments{The listed offsets give the position of the slit center
relative to the location of the target peakup.  The offsets are
oriented along P.A. = 112\arcdeg, and positive offsets correspond to
slit positions located east of the nucleus, as shown in Figure
\ref{figimages}.}

\end{deluxetable}
\end{center}

\begin{center}
\begin{deluxetable}{ccccccccc}
\tablewidth{7.2in} \tablecaption{Comparison of Models with Different
Features
\label{table2}}

\tablehead{\colhead{Surface} & \colhead{PSF} & \colhead{Wavelength} &
\colhead{Subsampling} & \colhead{\mbh} & \colhead{\ml} &
\colhead{\inc} & \colhead{$\theta$} & \colhead{\chisq} \\

\colhead{Brightness} & & \colhead{Shifts?} & \colhead{Factor} &
\colhead{(\msun)} & \colhead{(\mlsun)} & \colhead{(degrees)} &
\colhead{(degrees)} &}

\startdata

Constant & $\delta(r)$ & No & 2 & $1.18\times10^8$ & 4.03 & 63
& 27 & 143.6 \\

Exponential & $\delta(r)$ & No & 2 & 0.0 & 4.55 & 63 & 26 & 181.8 \\

Exponential & TinyTim & No & 1 & $1.32\times10^8$ & 4.21 & 62 & 26 &
142.2 \\

Exponential & TinyTim & Yes & 1 & $1.49\times10^8$ & 4.10 & 63 & 26 &
145.7 \\

Exponential & TinyTim & Yes & 2 & $8.12\times10^7$ & 4.34 & 64 & 26 &
181.3 \\

Exponential & TinyTim & Yes & 4 & $4.89\times10^7$ & 4.61 & 65 &
26 &  207.0 \\

WFPC2 & TinyTim & No & 2 & $2.04\times10^8$ & 3.79 & 63 & 27 & 115.9 \\

WFPC2 & TinyTim & Yes & 2 & $2.09\times10^8$ & 3.74 & 63 & 27 & 102.3 \\

\enddata

\tablecomments{All models were calculated for \rfit\ = 0\farcs5, with
no asymmetric drift correction, and with the four parameters \mbh,
\ml, \inc, and $\theta$ allowed to float in the model fits.  For the
\hn\ surface brightness, ``Exponential'' refers to a parameterization
of the form $S_0 + S_1 e^{-r/r_1}$, and ``WFPC2'' refers to the
surface brightness taken from the deconvolved, continuum-subtracted
WFPC2 F658N image.  The PSF was taken to be either a delta function at
the limit of the model resolution, or a STIS PSF generated by the Tiny
Tim program.  The entry ``Wavelength Shifts'' indicates whether the
models include instrumental shifts in wavelength for light entering
the slit off-center.  The last entry in the table is for a model with
the real \hn\ profile, the Tiny Tim PSF, and instrumental wavelength
shifts included; note that it results in the lowest value of \chisq\
by a significant margin.}

\end{deluxetable}
\end{center}

\clearpage

\clearpage

\clearpage

\clearpage

\clearpage

\clearpage

\clearpage

\clearpage

\clearpage

\clearpage

\clearpage

\clearpage 

\clearpage

\clearpage

\clearpage
 
\clearpage

\clearpage

\clearpage

\clearpage

\clearpage

\end{document}